\documentclass[aps,bibnotes,twocolumn,showpacs, showkeys, superscriptaddress,preprintnumbers,amsmath,amssymb,lengthcheck,prx]{revtex4-1}

\usepackage{graphicx}
\usepackage{dcolumn}
\usepackage{bm}
\usepackage{color}

\begin{document}

\title{Observation of a Group of Dark Rogue Waves in a Telecommunication Optical Fiber}

\author{F. Baronio} 
\email{fabio.baronio@unibs.it}
\affiliation{INO CNR and Dipartimento di Ingegneria dell'Informazione, Universit\`a di Brescia, Via Branze 38, 25123 Brescia, Italy}

\author{B. Frisquet}
\affiliation{Laboratoire Interdisciplinaire Carnot de Bourgogne, UMR 6303 CNRS/Universit\'e  Bourgogne Franche-Comt\'e, 21078 Dijon, France}

\author{S. Chen}
\affiliation{Department of Physics, Southeast University, Nanjing 211189, China}

\author{G. Millot}
\affiliation{Laboratoire Interdisciplinaire Carnot de Bourgogne, UMR 6303 CNRS/Universit\'e  Bourgogne Franche-Comt\'e, 21078 Dijon, France}

\author{S. Wabnitz} 
\affiliation{INO CNR and Dipartimento di Ingegneria dell'Informazione, Universit\`a di Brescia, Via Branze 38, 25123 Brescia, Italy}

\author{B. Kibler}
\affiliation{Laboratoire Interdisciplinaire Carnot de Bourgogne, UMR 6303 CNRS/Universit\'e  Bourgogne Franche-Comt\'e, 21078 Dijon, France}

\begin{abstract}
Over the past decade, the rogue wave debate has stimulated the comparison of 
predictions and observations among different branches of wave physics, particularly between
 hydrodynamics and optics, in situations where analogous 
 dynamical behaviors can be identified, thanks to the use of common universal
 models.  Although the scalar nonlinear Schr\"odinger equation (NLSE) has constantly played a central role 
 for rogue wave investigations, moving beyond the standard NLSE model is relevant and 
 needful for describing more general classes of physical systems and applications.  
 In this direction, the coupled NLSEs are known to play a pivotal role for the understanding of the complex wave dynamics 
 in hydrodynamics and optics.
 Benefiting from the advanced technology of high-speed telecommunication-grade components, 
 and relying on a careful design of the nonlinear propagation of orthogonally-polarized optical 
 pump waves in a randomly birefringent telecom fiber, this work explores, both theoretically and experimentally, 
 the rogue wave dynamics governed by such 
 coupled NLSEs. We report, for the first time, 
 the evidence of a group of three dark rogue waves, the so-called dark \textit{three-sister} rogue waves,
 where experiments, numerics, and analytics show a very good consistency.

\end{abstract}

													


\maketitle

\section{INTRODUCTION}
Rogue waves are large, unexpected and rare localized surface water waves which
can be extremely dangerous, even to large ships such as ocean liners \cite{perkin06}. Rogue waves present
a considerable hazard, because they are  unpredictable, may appear suddenly without any warning,
and can impact with tremendous force \cite{pelinosky08,kharif09}. The research on rogue waves in the ocean has been rapidly
developing over the past twenty years, and has also attracted the attention of many other areas of physics:
optics \cite{solli07,ak09,kibler10}, plasmas \cite{bailung11,bailung16,tsai16}, Bose-Einstein condensates \cite{bludov09,hulet17}, atmosphere \cite{stenflo09}, superfluid helium \cite{ganshin08,efimov10}.
Despite the versatile horizons of research in so many different physical domains, there is still an intense debate on the rogue wave generation mechanisms, 
which are closely related with the possibility to provide reliable predictions in engineering applications, such as the controllability of laser light emission, 
or even their suppression in the ocean. 
This rogue wave debate has stimulated the comparison of predictions and observations among distinct topical areas, 
in particular between hydrodynamics and optics, in situations where analogous dynamical behaviors can 
be identified, thanks to the use of common physical models \cite{on16,chen17review}.

The scalar nonlinear Schr\"odinger equation (NLSE) plays a central role for the study of rogue waves, as it provides an universal model
for wave evolution in hydrodynamics and optics. In deep water, in the absence of an ambient current, for single peaked or unimodal 
spectral conditions, the dynamics of water waves is well described by the NLSE \cite{on01}, which can be derived from
the Euler equations by assuming weak nonlinearity and narrow spectra. In optics, in singlemode optical fibers,
the temporal propagation of optical pulses is properly modeled by the NLSE \cite{agrabook}, which is derived from the Maxwell's 
equations by assuming a weak cubic (or Kerr) nonlinearity and under slowly varying envelope approximation (i.e., for narrow spectra relative to the carrier frequency).
Many properties of rogue wave dynamics can be described almost 
exactly by means of the analytic nonlinear solutions of the NLSE. These solutions represent the nonlinear stage of evolution of the modulation instability (MI) of plane waves,
namely the  Peregrine solitons \cite{pe83}, Akhmediev breathers  \cite{nail97}, and higher-order breather solutions on
a finite background \cite{nail97,nail09,ak13,kib15}. Some of these NLSE solutions are sometimes referred to as 
higher-order rogue waves or multi-rogue wave solutions. 
%
%
%
Experiments in water tanks and in singlemode fibers have successfully demonstrated the existence and the properties 
of the previously mentioned analytic solutions of the NLSE \cite{amin11,amin12,amin13,frisq13,frisq14,chab14}.
In particular, recent chaotic MI experiments have confirmed the existence of fundamental and higher-order periodic breathers \cite{narni16}. These observations are
in good quantitative agreement with numerical simulations based on the NLSE, 
starting from a fully noise-perturbed plane wave. Moreover, the rogue wave events found in the heavy-tailed 
statistics associated to noise-driven MI processes have been appropriately identified as collisions of 
elementary breathers (i.e., higher-order breather solutions) \cite{nail09,frisq13,narni16}. 

Although studies on rogue wave solutions of the NLSE are blooming in different areas of physics, moving beyond the
standard NLSE model is relevant and needful for describing more general classes of physical systems 
and applications. In this direction, recent studies have extended the search for rogue waves to coupled wave 
systems  \cite{baronio12,chen13,baronio14,chen14R,chen15,chen16,baro17}. In fact, a variety of physical phenomena 
require the modeling of waves with two or more components, in order to account for different spectral peaks, modes, 
or polarization states. 
Such coupled wave systems are of physical relevance in a multidisciplinary context, e.g., hydrodynamics, nonlinear optics, plasma physics, 
multicomponent Bose-Einstein condensates, and financial systems. 
More generally, multicomponent dynamical systems are usually occurring in the modeling of natural and human activities. 
Therefore, the analysis of such complex systems requires a study of the interactions between their multiple wave 
components.

In this context, the coupled NLSEs (vector NLSE), play a pivotal role for our understanding of rogue wave 
dynamics in hydrodynamics and optics. Oftentimes in oceanography the frequency spectra are characterized by the coexistence of two wave systems with 
different directions of propagation \cite{roske76}. This condition is known as a crossing sea: it implies that 
wave energy is concentrated over two different spectral peaks, and wave propagation is well described in terms of two coupled NLSEs \cite{miguel06,stenflo06,miguel11}. 
On the other hand, in optics it is well known that singlemode optical fibers are in fact bimodal. This is due to the unavoidable presence of 
residual birefringence, which may strongly influence the evolution of temporal pulse dynamics in optical fibers. 
Nonlinear pulse propagation in birefringent, singlemode optical fibers is properly modeled by a set of two coupled NLSEs \cite{menyuk87,menyuk96}. 
When compared with scalar NLSE dynamical systems, coupled NLSE systems may allow for cross-modulation and energy transfer 
between the different coupled components, that leads to a rich variety of rogue-wave dynamics.
%
%
%
%

%
%
%

In this work, we consider two orthogonally polarized optical pumps and explore their propagation dynamics in the normal dispersion regime of a randomly
birefringent telecom fiber. Whenever the degree of birefringence [or polarization mode dispersion (PMD)] is relatively low, 
wave propagation may be theoretically modeled by two coupled NLSEs, or Manakov system \cite{mana74}.
Recent studies of wave propagation in randomly birefringent and low PMD telecom fibers have demonstrated the emergence of complex 
polarization MI dynamics \cite{frisq15}, and the generation of dark rogue waves \cite{frisq16}. All of these phenomena can be
quantitatively described by using analytic solutions of the coupled NLSEs \cite{baronio14}. In the initial stage of the evolution of polarization MI, 
spectral sidebands experience an exponential amplification with distance, at the expense of the input pump waves  \cite{frisq15}.
Next, the nonlinear stage of MI may lead to the formation of dark rogue waves \cite{frisq16}. Similar to the Fermi-Pasta-Ulam recurrence behavior of the scalar NLSE, the evolution of coupled wave solutions of the Manakov system is expected to display a cyclic energy exchange between multiple spectral modes, which may result in the emergence of a complex dark vector rogue wave dynamics. 

Here we experimentally unveil, to our knowledge for the first time, the emergence of nonlinear wave recurrence phenomena in the vector rogue wave dynamics. 
Namely, we demonstrate the existence of complex nonlinear superpositions of the coupled rogue waves, forming a group of three dark rogue waves (often dubbed as \textit{three-sister} rogue waves) which are localized in both time and space. Such class of higher-order rogue wave solutions represent the simplest nonlinear prototype of MI-driven rogue waves in multi-component nonlinear wave systems.

This work is organized as follows. In Sec. II, we provide the theoretical framework describing dark \textit{three-sister} rogue waves based on coupled NLSEs of Manakov type. In Sec. III we describe our experimental setup for observing such localized structures based on the excitation of polarization MIs. Next, in Sec. IV we compare theoretical and numerical predictions with experimental demonstrations. Lastly in Sec. V, we briefly discuss the relevance of our findings, and highlight the perspectives for further work.

\section{THEORETICAL MODEL}
The propagation of two orthogonally polarized optical pump waves in 
a random weakly birefringent telecom fiber
is described by the two coupled NLSEs  \cite{mana74,frisq15}: 
\begin{eqnarray}\label{3wridim}
\nonumber i  \frac{\partial E_1}{\partial z} +i  \beta^{'}_1 \frac{\partial E_1}{\partial t}-  \frac{\beta^{''}_1 }{2}  \frac{\partial^2 E_1}{\partial t^2} +\gamma_1  (|E_{1}|^2+|E_{2}|^2) E_1&=& 0,\\ 
i \frac{\partial E_2}{\partial z} +i   \beta^{'}_2 \frac{\partial E_2}{\partial t}-  \frac{\beta^{''}_2}{2}  \frac{\partial^2 E_2}{\partial t^2} +\gamma_2 (|E_{1}|^2+|E_{2}|^2) E_2&=& 0.
\end{eqnarray}
where $E_{1,2} (W^{1/2})$ is the complex slowly varying envelope of the pump waves at frequency $\omega_{1,2}$, respectively; 
$\beta^{'}_{1,2} (s m^{-1}) $  is the inverse group velocity, $\beta^{''}_{1,2} (s^2 m^{-1}) $ is the group-velocity dispersion, and $\gamma_{1,2} (W^{-1}m^{-1})$  is the effective Kerr 
nonlinear coefficient. $z (m)$ denotes propagation distance,  and $t (s)$ is time.  
We set $\omega_{1,2}=\omega_0 \pm \frac{1}{2} \Delta \omega$, where $\Delta \omega= 2 \pi \Delta f$ is the relative frequency 
offset of the two waves. For optical fibers, the small difference between the group velocity dispersion and 
nonlinear coefficient of the two polarization components may be neglected for a typical small frequency offset $\Delta f$. Thus we may set  $\beta^{''}_{1,2}= \beta^{''}$, $\gamma_{1,2}=\gamma$, i.e., the dispersion and nonlinearity are evaluated
at the mid-frequency $\omega_0$. The nonlinear coefficient $\gamma= \frac{8 n_2 \omega_0} {9cA_{e}}$, with $c$ the vacuum light velocity, 
$n_2=2.6 \times 10^{-20} m^2 W^{-1}$ is the nonlinear Kerr index coefficient, and $A_{e} (m^2)$ is the effective area of propagating modes
(equal for the two polarization modes).  
The power $P=P_1+P_2=|E_1|^2+|E_2|^2 (W)$.

Note that the NLSEs (\ref{3wridim}) are incoherently coupled (i.e., without coherent coupling terms), with self-phase and cross-phase modulation coefficients both equal to unity. This condition is well-satisfied when the two orthogonally polarized components of a vector field propagating in telecom optical fibers exhibit a rapidly-varying weak birefringence, which is the typical situation in long-haul optical communication links. Moreover, higher-order dispersion, material absorption and Raman effects can be safely neglected in our experimental conditions (for more details, see Refs. \cite{menyuk87,menyuk96,frisq15}).

With the transformations $u_{1,2} =E_{1,2} / \sqrt{P_0}, $ $\tau=(t-\frac{z} {v_{g}})\frac{1} {t_{0}}$, $\xi=z/z_0$,
where $z_0=1/ (\gamma P_0)$, $t_0=\sqrt{\frac{\beta^{''}} {\gamma P_0}}$, $\delta=\frac{\beta^{'}_{1}-\beta^{'}_{2} } {2 \beta^{''}} t_0 $ ($\delta= \pi \Delta f t_0$),
$v_g=\frac{2} {\beta^{'}_{1}+\beta^{'}_{2} }$, we reduce Eqs. (\ref{3wridim}) to the dimensionless form \cite{menyuk87} of the Manakov system
%
%
\begin{eqnarray}\label{3wri}
\nonumber i  \frac{\partial u_1}{\partial \xi} +i  \delta \frac{\partial u_1}{\partial \tau}-  \frac{1}{2}  \frac{\partial^2 u_1}{\partial \tau^2} + (|u_{1}|^2+|u_{2}|^2) u_1&=& 0,\\ 
i \frac{\partial u_2}{\partial \xi} -i  \delta \frac{\partial u_2}{\partial \tau}-  \frac{1}{2}  \frac{\partial^2 u_2}{\partial \tau^2} + (|u_{1}|^2+|u_{2}|^2) u_2&=& 0.
\end{eqnarray}
Equations (\ref{3wri}) exhibit multi-component dark rogue soliton solutions, localized both in space and in time, 
with the property of describing unique wave events \cite{baronio14,chen14}. Here we consider the case of equal amplitude optical pump waves. 
In this situation, the fundamental dark rogue waves can be expressed as  \cite{baronio14,chen14}
\begin{equation} 
u_{j}^{[1]}=u_{j0}\left[1-\frac{3i\Lambda\theta_{j}^{*}\vartheta/\alpha_{j}^{*}}
{|\vartheta|^{2}-a^2(|\theta_{1}/\alpha_{1}|^{2}+|\theta_{2}/\alpha_{2}|^{2})}\right], \label{fun}
\end{equation}
where $j=1,2$ and the same below,
\begin{gather*}\label{def0}
u_{j0} =ae^{-i\left(v_{j}\tau-k_{j}\xi\right)}e^{-i[(-1)^{j}\delta\tau+\frac{\delta^{2}}{2}\xi]},  \\ k_{j} =2a^{2}+v_{j}^{2}/2, ~~~
\kappa=v_{1}+v_{2},~~~\upsilon=v_{1}-v_{2}, \\
\eta =\pm \left[4\sqrt{a^{2}(a^{2}+\upsilon^{2})}-4a^{2}-\upsilon^{2}\right]^{1/2}, \\
\vartheta=\tau-(\kappa+i\eta)\xi/2, \ \ \ \alpha_{j}=(\kappa+i\eta)/2-v_{j}, \\
\theta_{j}=\vartheta-i/\alpha_{j},  \ \ \Lambda = (8a^{2}-3\eta^{2}-\upsilon^{2})/(6\eta).
\end{gather*}

%
The evolution of the fundamental dark rogue soliton solution (\ref{fun})  
depends on the dimensionless parameters $a$, $v_1$ and $v_2$, which originate from the 
initial optical pumps (namely their amplitudes, and their relative frequency shift).
Such a dark solution is localized in both $\xi$ and $\tau$, thus it describes a unique dark hole wave event. It is written in terms
of rational functions, in contrast to propagation-invariant dark solitons of the coupled NLSEs  \cite{kiv93}.
The existence of this elementary coupled dark rogue solution has been recently demonstrated in Ref. \cite{frisq16}.
However the general solution for space and time localized rogue waves has a hierarchy of increasing
order. 
%
For the coupled NLSEs, the second-order dark rogue wave solutions has the expression  \cite{chen14}
\begin{equation}
u_{j}^{[2]}=u_{j0}\left[1+\frac{3\Lambda(R_{j}^{*}Q+S_{j}^{*}W)}{a(M_{11}M_{22}-M_{12}M_{21})}
\right], \label{sec}
\end{equation}
%
%
%
where 
\begin{gather*}\label{def}
Q=R_{0}M_{22}-S_{0}M_{21}, \ \ \ W=S_{0}M_{11}-R_{0}M_{12},   \\
R_{0}=2\varepsilon_{1}-2i\varepsilon_{2}\phi\vartheta,  \ \ \ R_{j}=2a(i\varepsilon_{1}+\varepsilon_{2}\phi \theta_{j})/\alpha_{j}, \\
S_{0}=\varepsilon_{1}p-i\varepsilon_{2}\phi(q\vartheta-2\sigma\xi)+2\varepsilon_{3}-2i\varepsilon_{4}\phi\vartheta,  \\
S_{j}=\frac{a}{\alpha_{j}^{2}}\left[iS_{0}\alpha_{j}-2\varepsilon_{1}(\phi^{2}\theta_{j}+i\sigma)
-2i\varepsilon_{4}\phi\right]+ \\
\ \ \ \ \ -\frac{a\varepsilon_{2}\phi}{\alpha_{j}^{3}}\left[(iq+2\sigma\vartheta-2i\phi^{2}\vartheta^{2}/3)\alpha_{j}-2(
\phi^{2}\theta_{j}+2i\sigma)\right],  \\
M_{11}=|R_{0}|^{2}-|R_{1}|^{2}-|R_{2}|^{2}, \\
M_{12}=R_{0}^{*}S_{0}-R_{1}^{*}S_{1}-R_{2}^{*}S_{2}-M_{11}\equiv M_{21}^{*}, \\
M_{22}=|S_{0}|^{2}-|S_{1}|^{2}-|S_{2}|^{2}-M_{12}-M_{21}, \\
K = (8a^{2}+3\eta^{2}-\upsilon^{2})/(4\eta), \\
\phi=\frac{1}{2}\sqrt{3\Lambda(\eta^2+\upsilon^2)/K}, \\
\varphi=\frac{i(3\Lambda^2+\phi^2)}{2K},~~~~~~ \sigma =\varphi -i\Lambda, \\
p= -\phi^{2}\vartheta^{2}-2i\varphi \vartheta+(i\eta\Lambda+2\varphi K) \xi, \\
q=p+\frac{2}{3}\phi^{2}\vartheta^{2}-i(\sigma-3\varphi)/K+\sigma^{2}/\phi^{2}.
\end{gather*}
where $\varepsilon_1, \varepsilon_2, \varepsilon_3, \varepsilon_4$ are four arbitrary complex constants.
The parametric existence condition for the above rogue solutions (\ref{fun}) and (\ref{sec}) is $|v|< 2 \sqrt{2}a$.

It is clear that, besides the parameters $a$, $v_1$ and $v_2$, which originate from the input (or naked) optical pumps,
the dynamics of the dark rogue wave solution (\ref{sec}) also depends on the complex constants $\varepsilon_1, \varepsilon_2, \varepsilon_3, \varepsilon_4$, which
rule the higher order rogue wave superpositions. Note that solution (\ref{sec}) reduces to the solution (\ref{fun})
when $\varepsilon_1 \neq 0$ and $\varepsilon_2=0$.

\section{EXPERIMENTAL SETUP}
\begin{figure}[h!]
\begin{center}
\includegraphics[width=6cm]{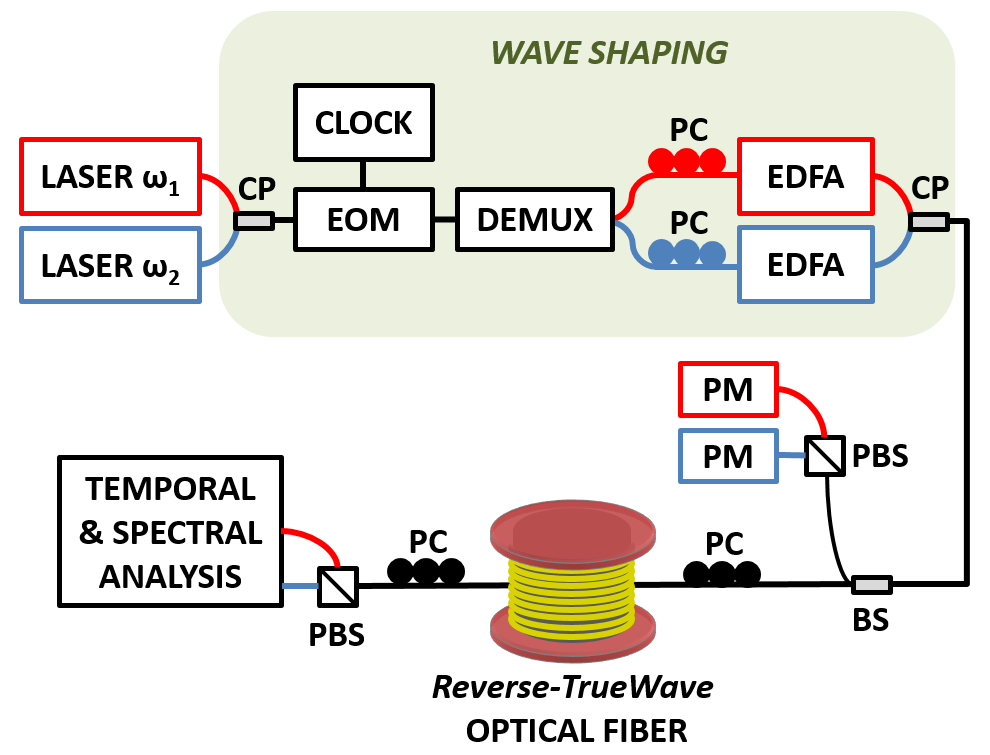}
    \end{center}
     \caption{Experimental setup: red and blue lines depict the two orthogonally polarized pump waves with distinct frequencies. CP: 50/50 fiber couplers; EDFA: Erbium doped fiber amplifier; EOM: electro-optic (intensity) modulator; PC: polarization controller; BS: 10/90 beam splitter; PBS: polarization beam splitter; PM: power-meter.
    } \label{fexp}
\end{figure}

For the generation and observation of the multicomponent dark rogue waves, we utilized the experimental configuration illustrated in  Fig. \ref{fexp}.
Two external-cavity diode lasers generate the pump waves, at frequencies  $\omega_1$ and $\omega_2$, which are fixed around the central frequency $\omega_0$ (with $\omega_0= 2 \pi c / \lambda_0$, and $\lambda_0 = 1554.7 nm$), which results in the frequency spacing $\Delta \omega= 2 \pi \Delta f$, with $\Delta f=100 GHz$. The two pumps are superimposed with a polarization maintaining fiber optical coupler with a 50:50 coupling ratio. We developed a careful linear wave shaping stage to excite polarization modulation instability at a particular frequency in the baseband regime, in order to satisfy the condition for the emergence of rogue wave solutions \cite{baronio14,frisq16}. To this end, we use an electro-optic modulator, driven by a $35-GHz$ RF clock, to induce the initial sinusoidal perturbations to the pumps. Note that, in order to reach the high pumping powers required in our experiments, we added another modulation stage (not shown in Fig.  \ref{fexp}) to create $100 ns$ square pulse trains with a duty cycle of 1:10 for both pump waves. Such pulses still provide a quasi-continuous wave background condition, since the pulse duration is more than three orders of magnitude longer than the excited MI period. Moreover, to suppress the detrimental stimulated Brillouin scattering (SBS) effect that may occur in the optical fiber, a supplementary phase modulation stage (not shown in Fig.  \ref{fexp}) was also inserted \cite{frisq15}. The two pump waves are spectrally separated by means of a programmable demultiplexer (DEMUX). A pair of polarization controllers allow us to manage two pumps with orthogonal linear states of polarization, amplified independently by erbium-doped fiber amplifiers, and finally recombined and injected into the fiber. The total input power was $P = 2.5 W$. 

The optical fiber used in our experiment is a reverse-TrueWave fiber with a chromatic dispersion of $-14 ps/nm/km$, a nonlinear coefficient  $\gamma=2.4W^{-1}km^{-1}$ and an attenuation of $0.25 dB/km$ at $\lambda_0$. This fiber has a very low PMD of $0.017 ps \, km^{-1/2}$. Two fiber spans were used in our experiments: a $3 km$-long span, and a $5 km$-long span. At the fiber output, a polarization beam splitter (PBS) selects the output light propagating in the two orthogonal linear polarizations. The output light is simultaneously analyzed both in the spectral and temporal domains by means of an optical spectrum analyzer (with $0.02 nm$ resolution bandwidth) and an optical sampling oscilloscope (with $0.8 ps$ resolution). Note that the orthogonal polarization states were continuously tracked before injection into the fiber, thanks to power measurements after their separation with another polarization beam splitter.

\section{RESULTS}

\begin{figure}[h!]
\begin{center}
\includegraphics[width=7.cm,height=3.8cm]{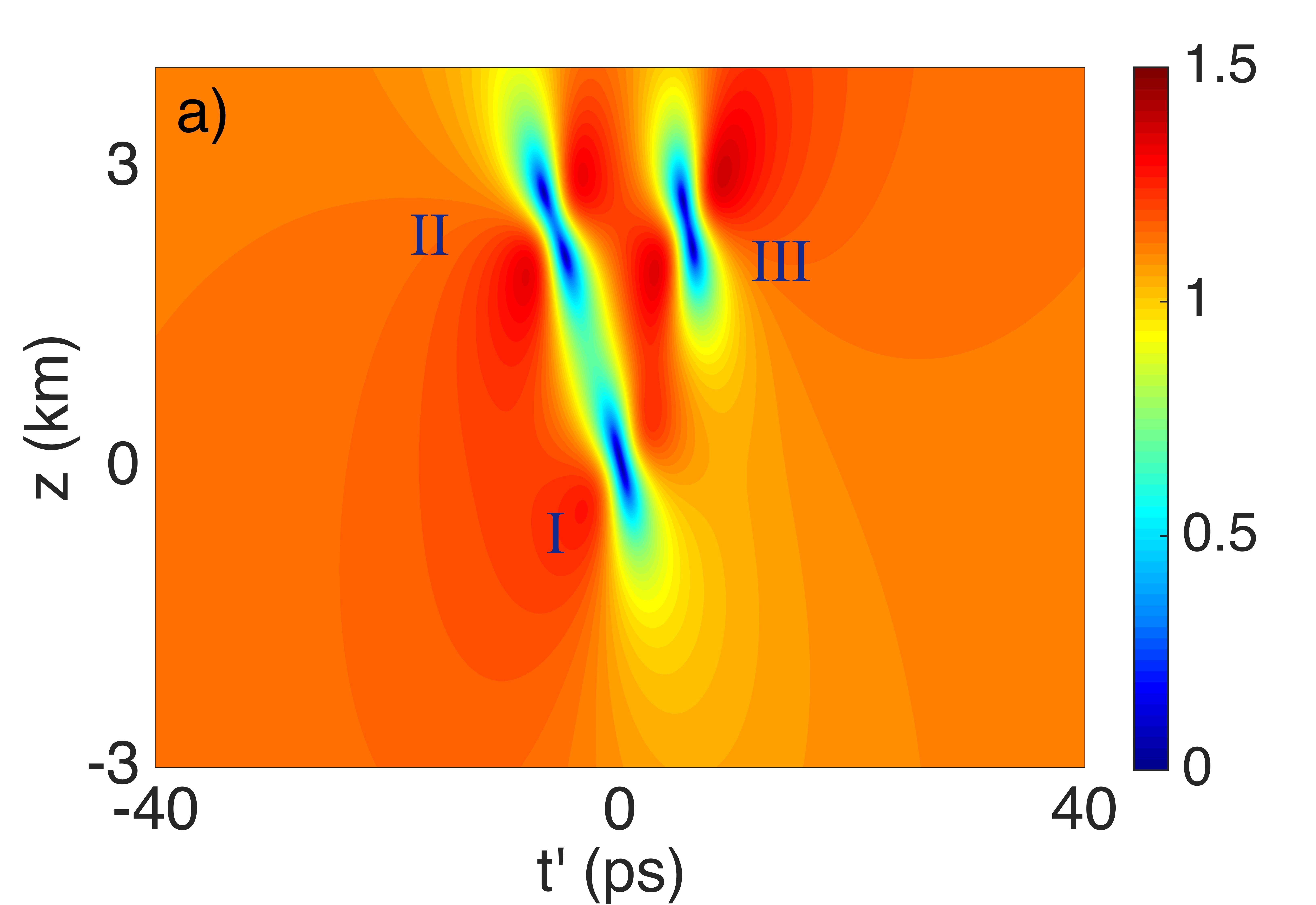}
\includegraphics[width=7.cm,height=3.8cm]{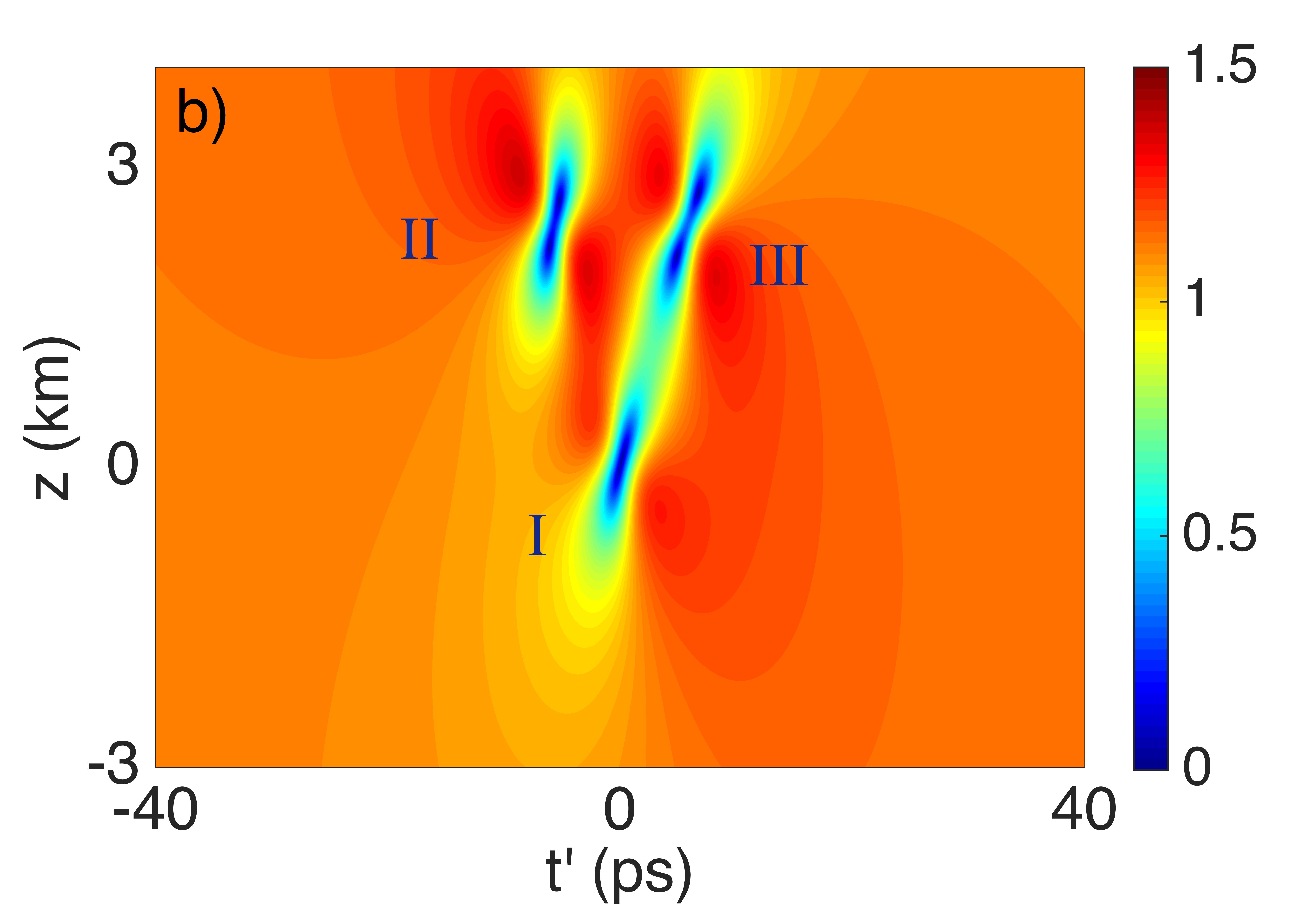}
    \end{center}
     \caption{Contour plot of the two orthogonal polarization waves  a) $|E_1(t',z)| $ and  b) $|E_2(t',z)|$, describing
     the analytical three dark rogue wave solution of the coupled NLSEs ($t'=t-{z} /{v_{g}}$). 
     Parameters $\beta^{''}=18ps^2/km$, $\gamma=2.4W^{-1}km^{-1}$, $P_{1,2}=1.25 W $, $\Delta f=100GHz$.
     Moreover: $a=1$, $v_1=-v_2=\delta=-0.77$,  $\eta=-0.986$; $\varepsilon_1=3.25 i, \varepsilon_2=-4.25 i, \varepsilon_3=7 i, \varepsilon_4=0$.  
    } \label{fig1}
\end{figure}

In order to properly set up our input conditions in accordance with the experimental parameters of the optical fiber, we present the analytical second-order dark rogue soliton solution of the coupled NLSEs (\ref{3wridim}).
Figure \ref{fig1} illustrates the spatio-temporal variation of $|E_1(t',z)|$ and $|E_2(t',z)|$ of a second-order rogue 
wave solution (\ref{sec}), in a reference frame traveling
with the average group velocity of the two pumps ($t'=t-{z} /{v_{g}}$). 
We have set $\beta^{''}=18ps^2/km$ (equivalent to $-14 ps/nm/km$), $\gamma=2.4W^{-1}km^{-1}$, $P_{1,2}=1.25 W (P=2.5 W)$, $\Delta f=100GHz$.
Moreover, $a=1$, $v_1=-v_2=\delta=-0.77$, and $\eta=-0.986$; $\varepsilon_1=3.25 i, \varepsilon_2=-4.25 i, \varepsilon_3=7 i, \varepsilon_4=0$ (we note that, when $v_1=-v_2=\delta$, there is no phase dependence of the rogue components (\ref{sec}) versus the temporal coordinate).
Note that Fig. \ref{fig1} shows a nonlinear superposition of three dark waves, which are localized both in time and in space
in a triangular configuration, dubbed as \textit{three-sister} dark rogue waves. 
Each field component of the first dark wave (I) possesses a zero-amplitude point, while the second (II) and third (III) 
dark structures possess two-zero-amplitude points, along with a central sink built on the finite continuous wave background.

\begin{figure}[h!]
\begin{center}
\includegraphics[width=7.cm,height=3.8cm]{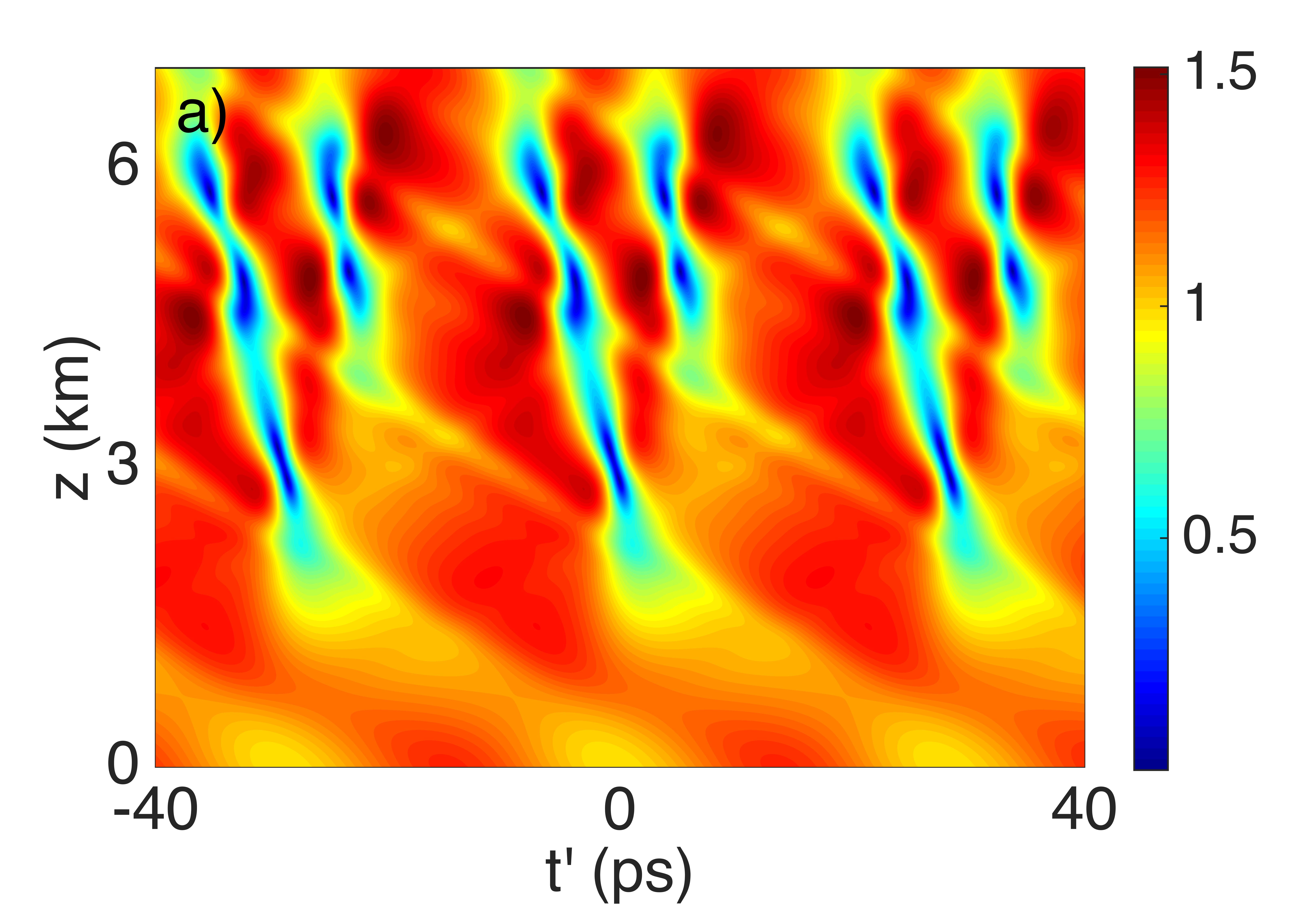}
\includegraphics[width=7.cm,height=3.8cm]{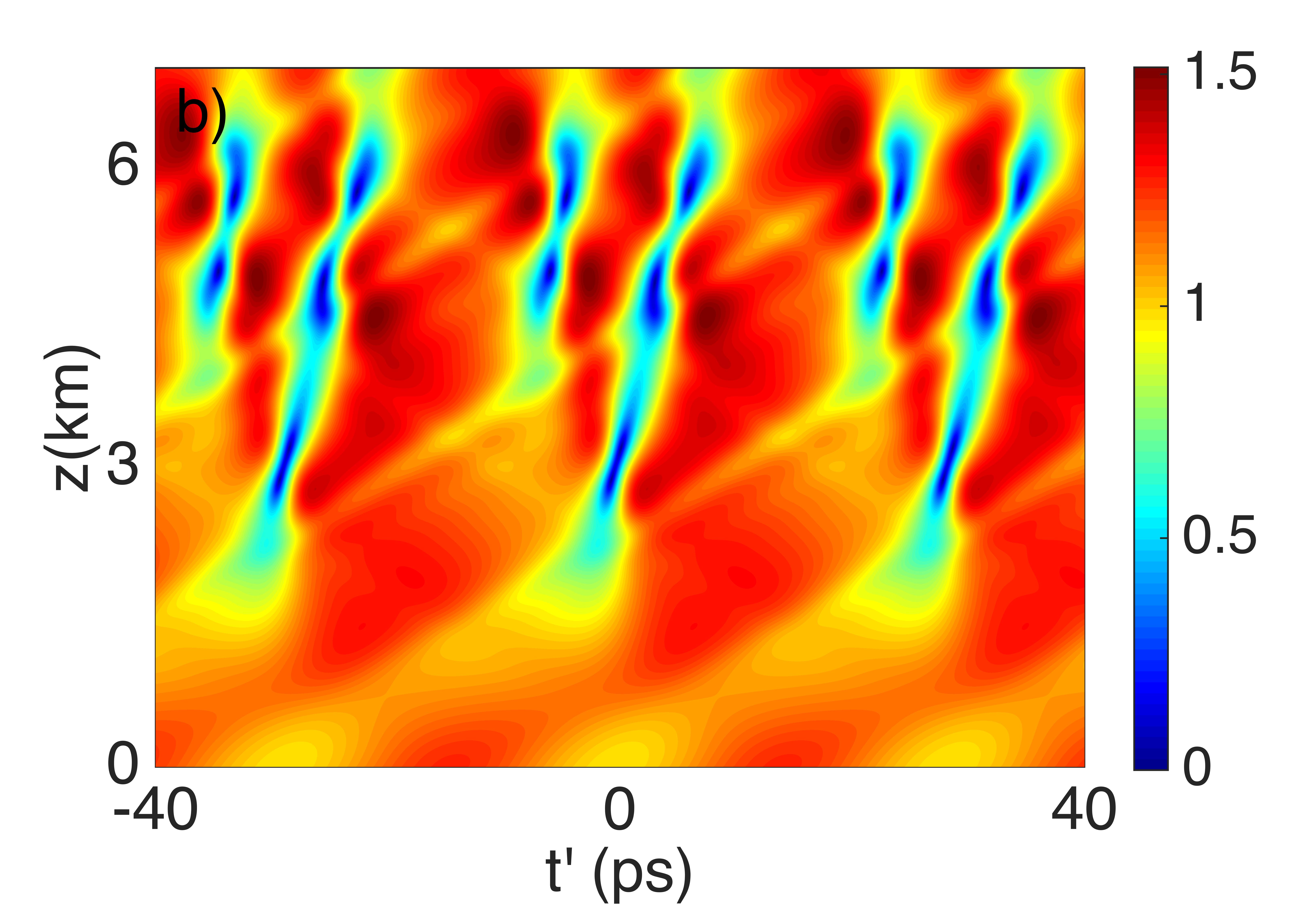}
    \end{center}
     \caption{Contour plot of the two orthogonal polarization waves  a) $|E_1(t',z)| $ and  b) $|E_2(t',z)|$, describing
     the numerical excitation of dark rogue waves. 
     Parameters $\beta^{''}=18ps^2/km$, $\gamma=2.4W^{-1}km^{-1}$, $P_{1,2}=1.25 W$, $\Delta f=100GHz$.  
     As to the modulating signal, $\epsilon=0.13$ and $f_{m}=35 GHz$.  } \label{fig2}
\end{figure}


Next, we numerically investigated the possibility to experimentally generate the \textit{three-sister} dark rogue waves, 
under realistic, but non-ideal, experimental conditions at the input of the optical fiber. To this end, we
characterized the propagation of the two continuous wave pumps injected into our fiber, when the modulating 
signal is present. Clearly, the use of a periodic modulation will result in the generation of time periodic dark wavetrains. 
We made use of a standard split-step Fourier technique, that is
commonly adopted in the numerical solution of the coupled NLSEs Eqs. (\ref{3wridim}).
As input signal in our simulation, we took the wave envelopes  
$E_{1,2}(t',z=0)=\sqrt{P_{1,2}}-\epsilon \, cos(2\pi f_{m} \, t')$, 
with $\epsilon=0.13$ and $f_{m}=35 GHz$
(we may note that, in our experimental setting, there is no phase dependence of injected waves versus 
the temporal coordinate). 
Figure \ref{fig2} shows the numerically computed $(t',z)$ dependence of the amplitude of solutions of the coupled NLSEs (\ref{3wridim}),
which describe the nonlinear dynamics of the two orthogonal optical pumps. As it can be seen, one observes the
generation of time-periodic notch structures, with are localized both in time and in space. 
Figure \ref{fig2} shows that each individual dark structure within the periodic wavetrain closely matches the shape of the
\textit{three-sister} dark rogue wave that was previously presented in Fig. \ref{fig1} 
(with a $3 km$ translation of the longitudinal evolution $z$ coordinate).

\begin{figure}[h!]
\begin{center}
\includegraphics[width=4cm,height=2.6cm]{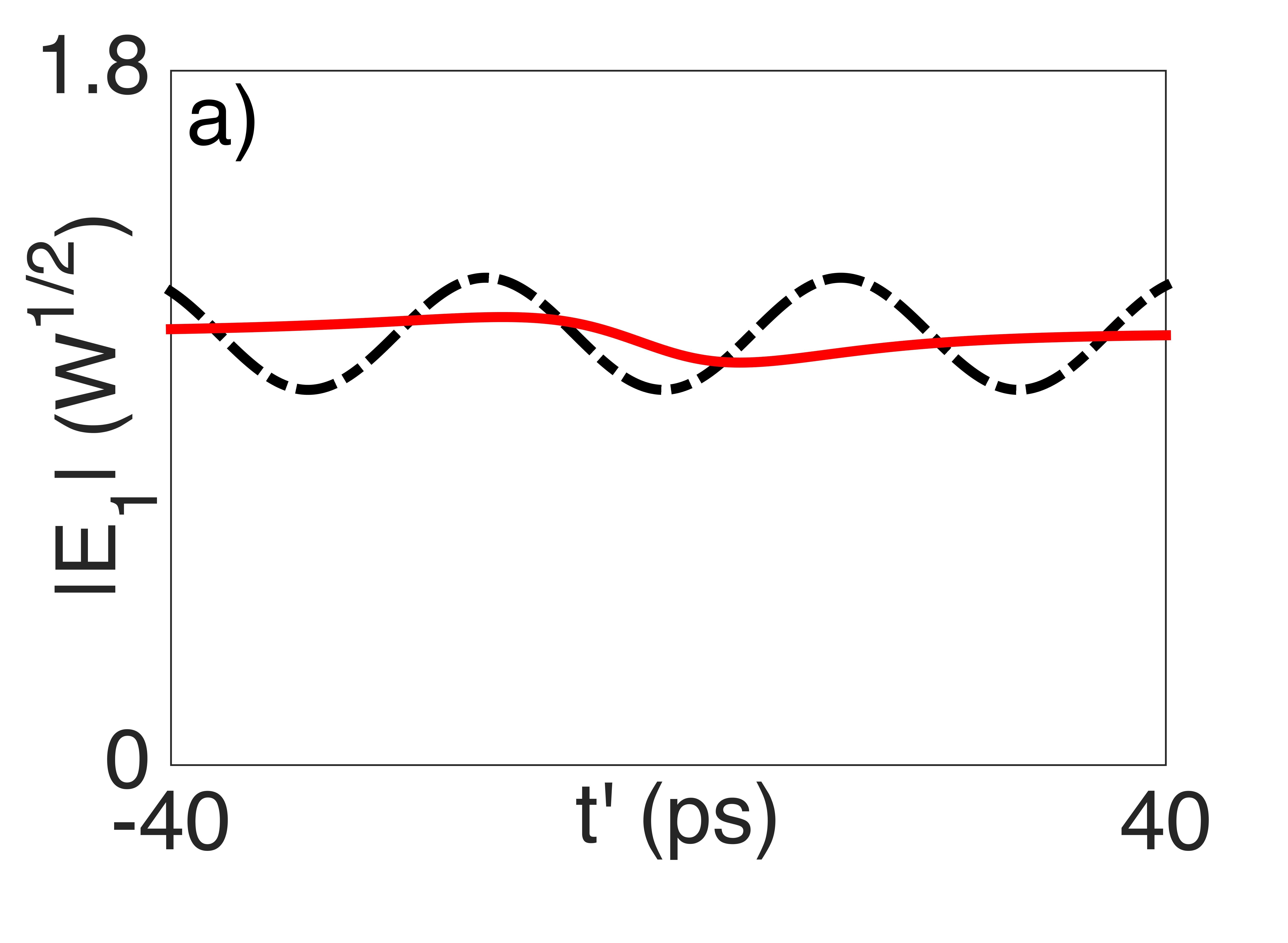}
\includegraphics[width=4cm,height=2.6cm]{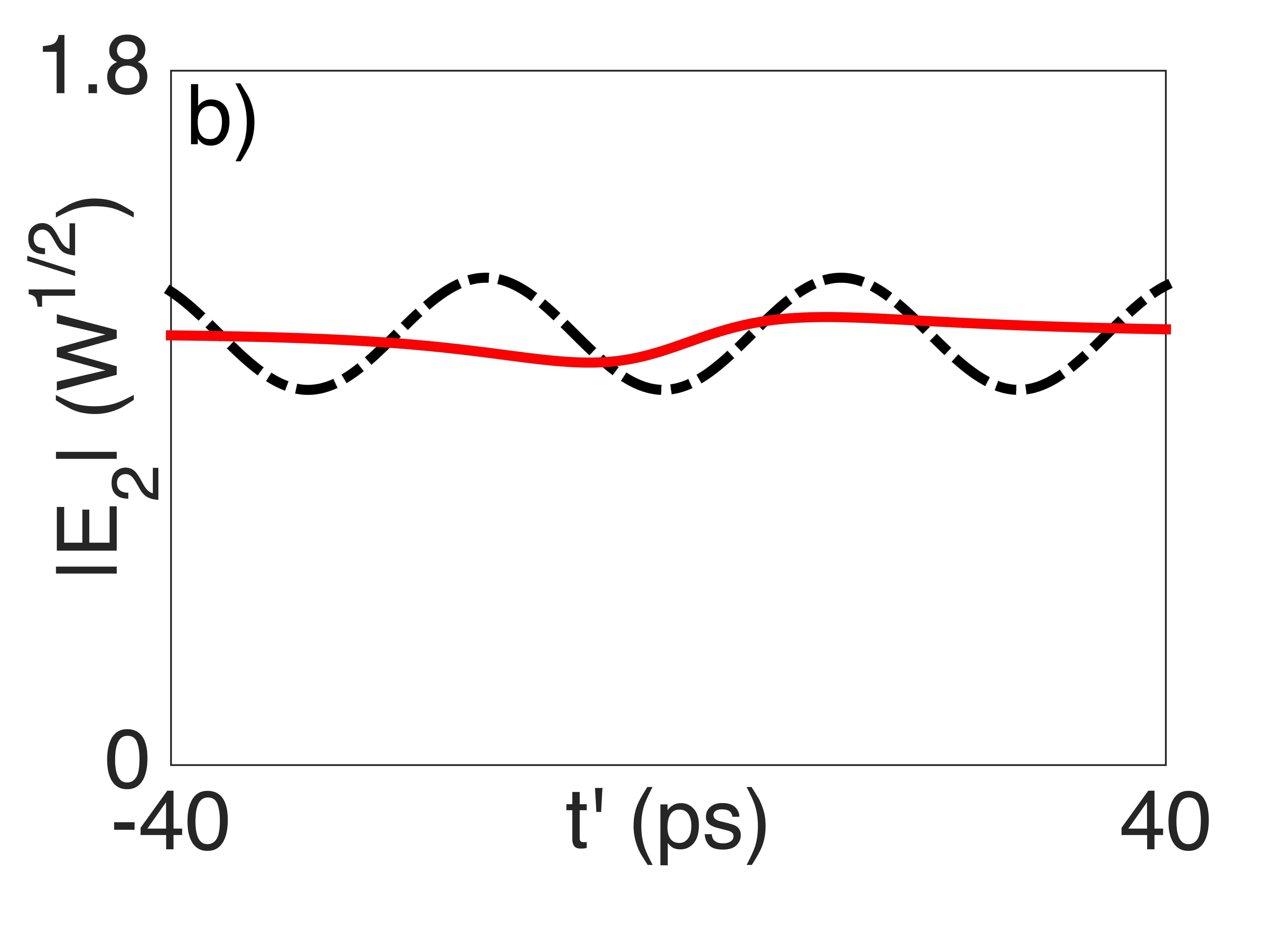}
\includegraphics[width=4cm,height=2.6cm]{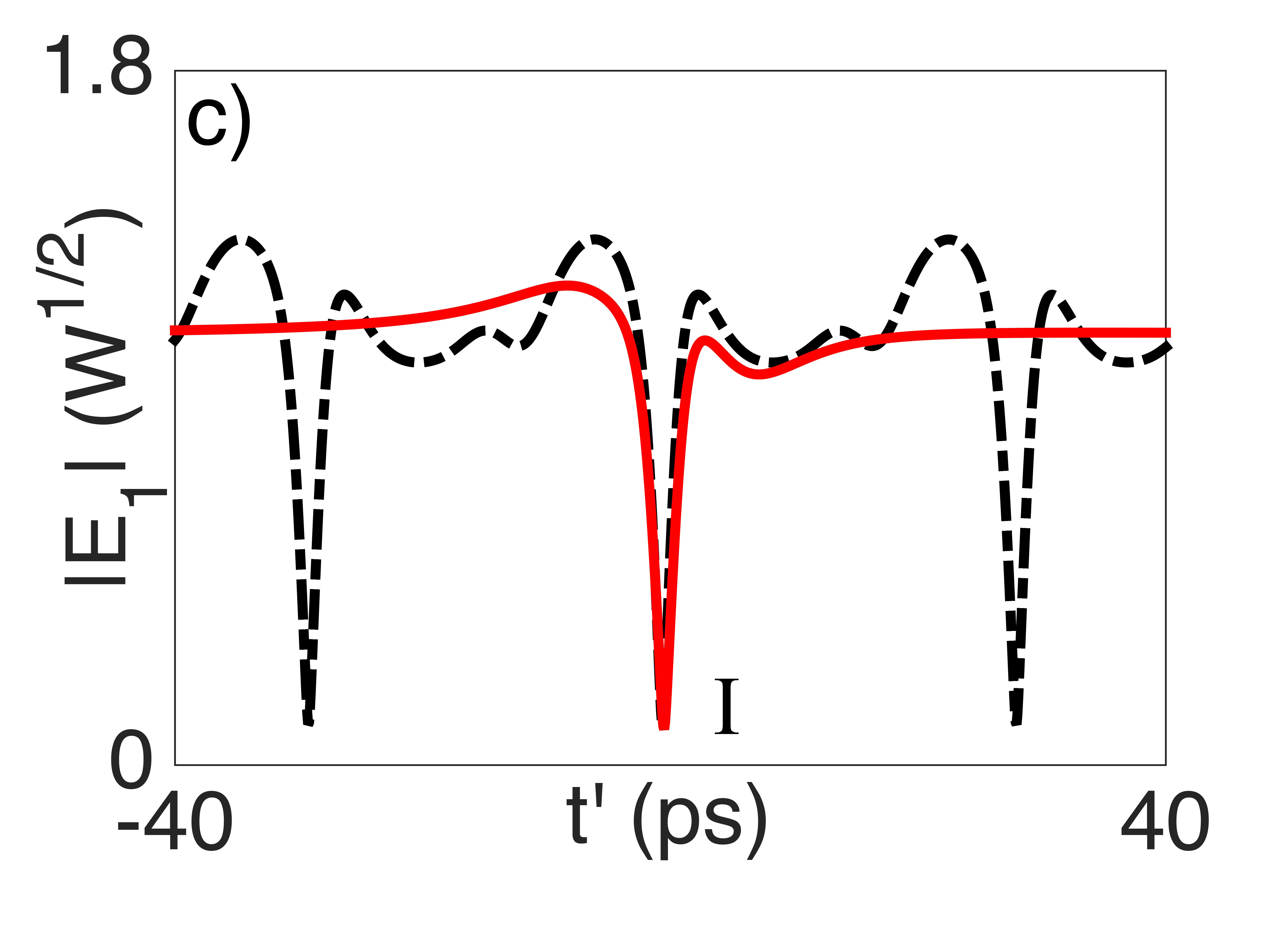}
\includegraphics[width=4cm,height=2.6cm]{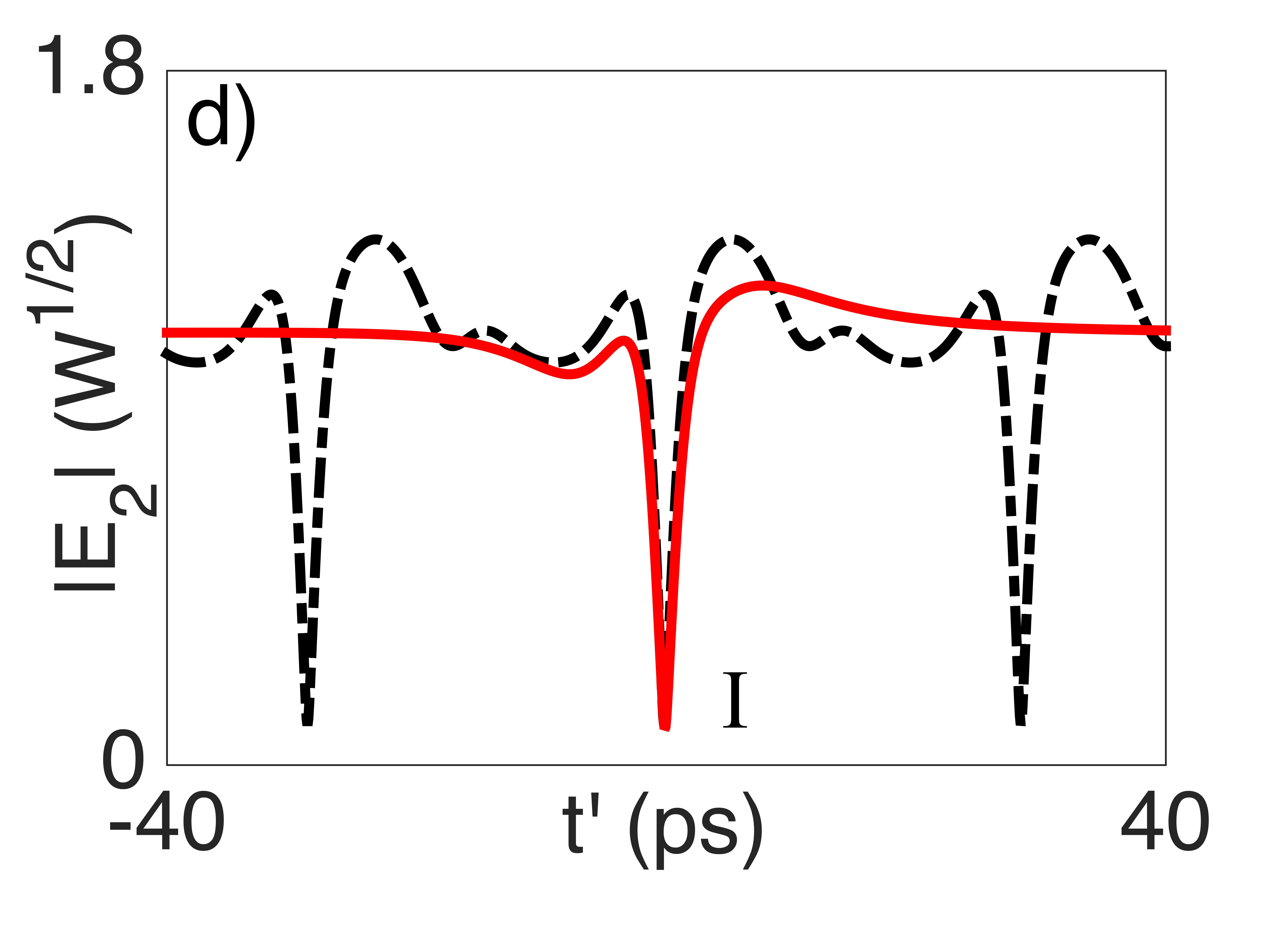}
\includegraphics[width=4cm,height=2.6cm]{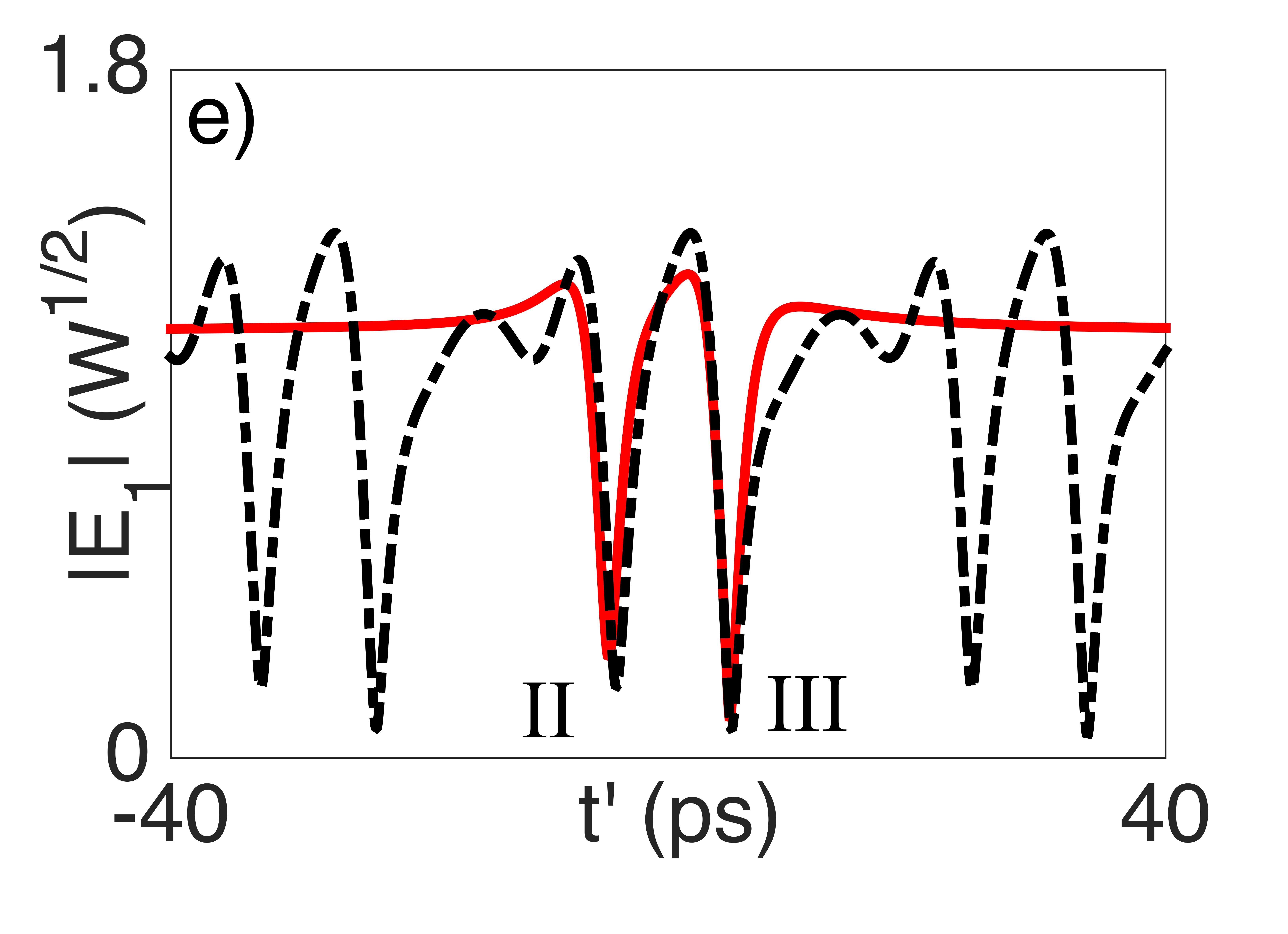}
\includegraphics[width=4cm,height=2.6cm]{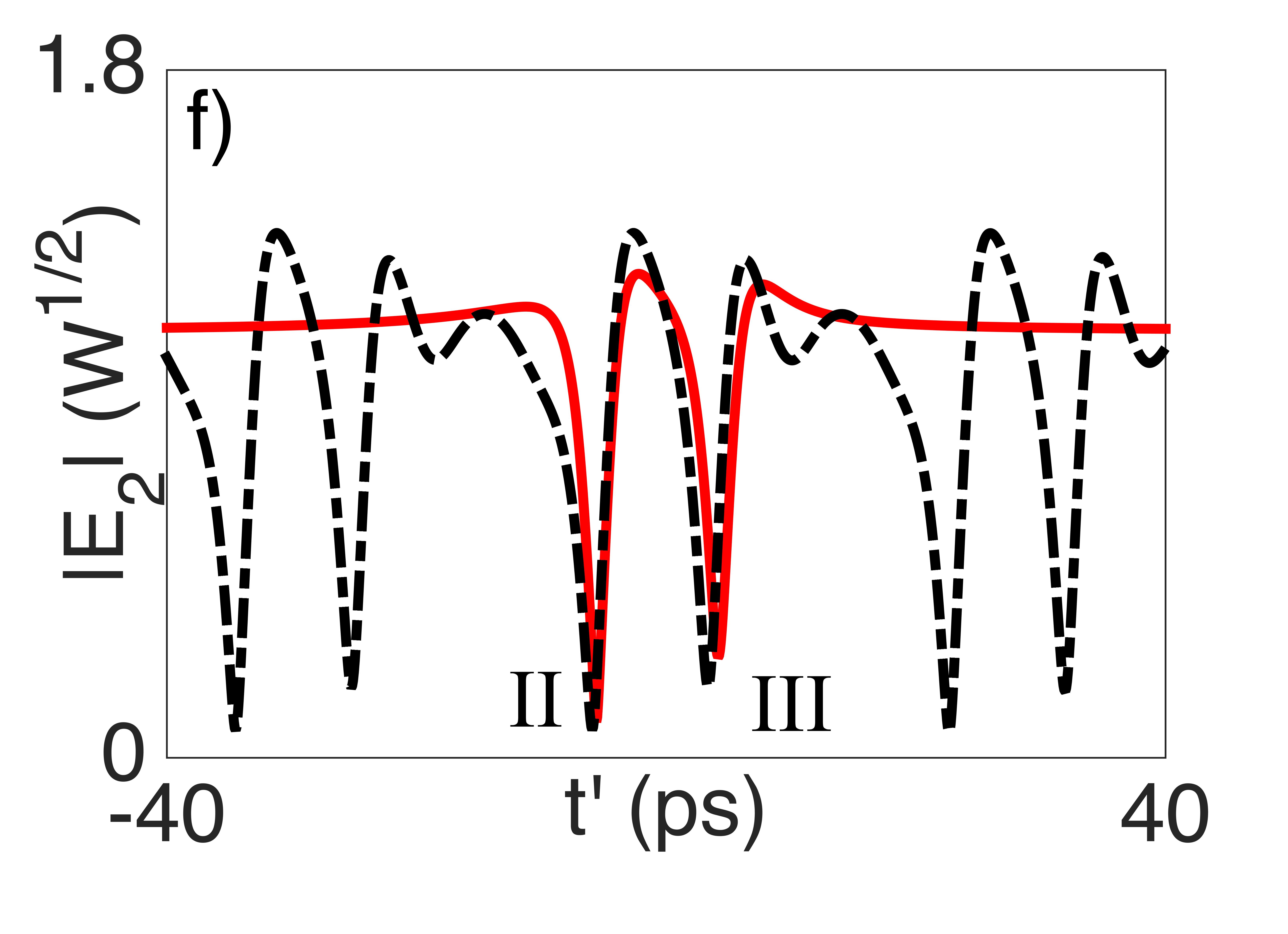}
    \end{center}
     \caption{Comparison between analyical (red line) and numerical (black dashed line) temporal profiles of 
     $|E_1(t')|$ and $|E_2(t')|$ at the input (a,b), after $3 km$ (c,d) and after $5 km$ (e,f). Theoretical 
     results refer to the case reported in figure \ref{fig1}, numerical ones to figure  \ref{fig2}.    
    } \label{fig3}
\end{figure}

 \begin{figure}[h!]
\begin{center}
\includegraphics[width=4cm,height=2.6cm]{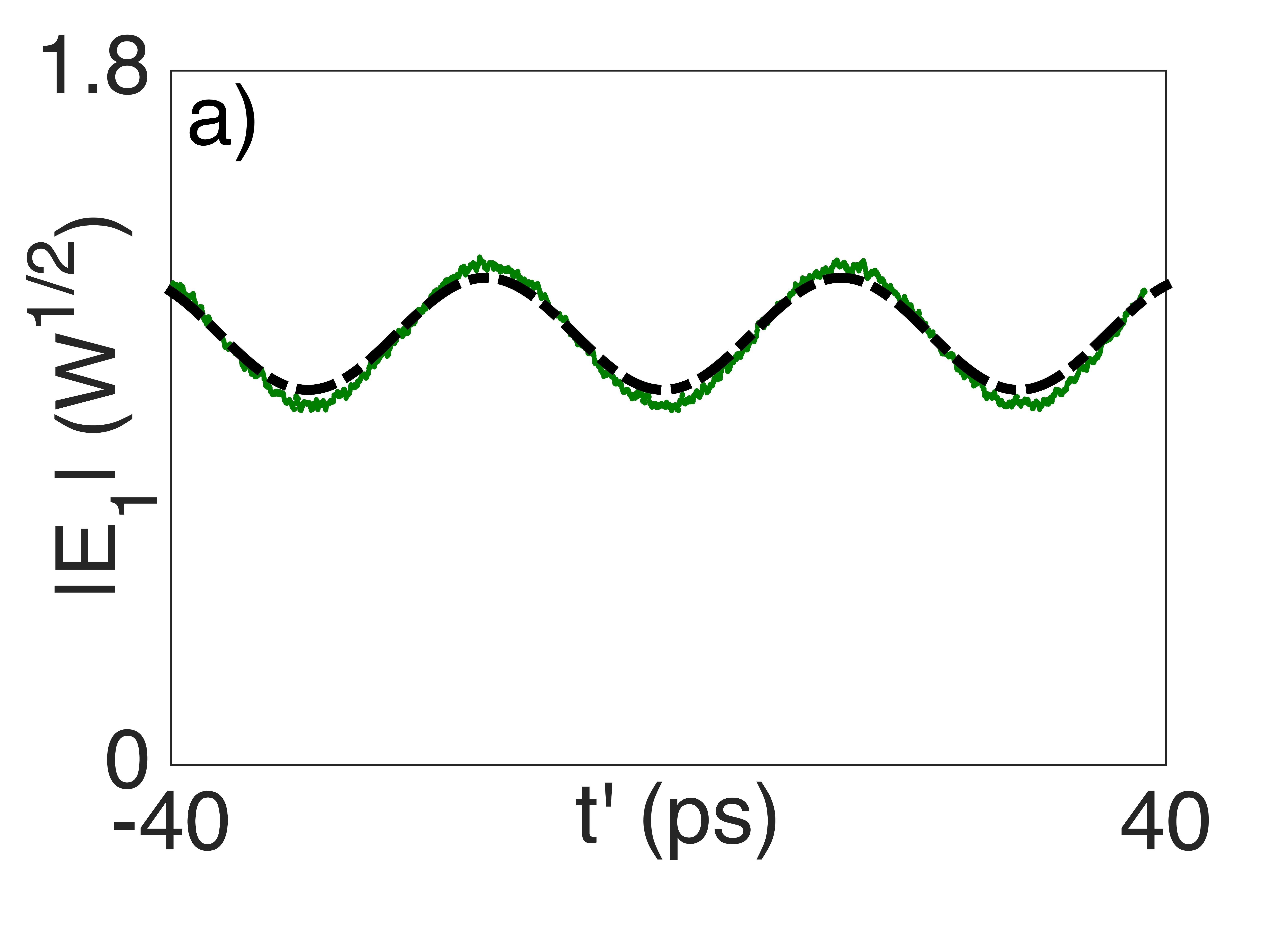}
\includegraphics[width=4cm,height=2.6cm]{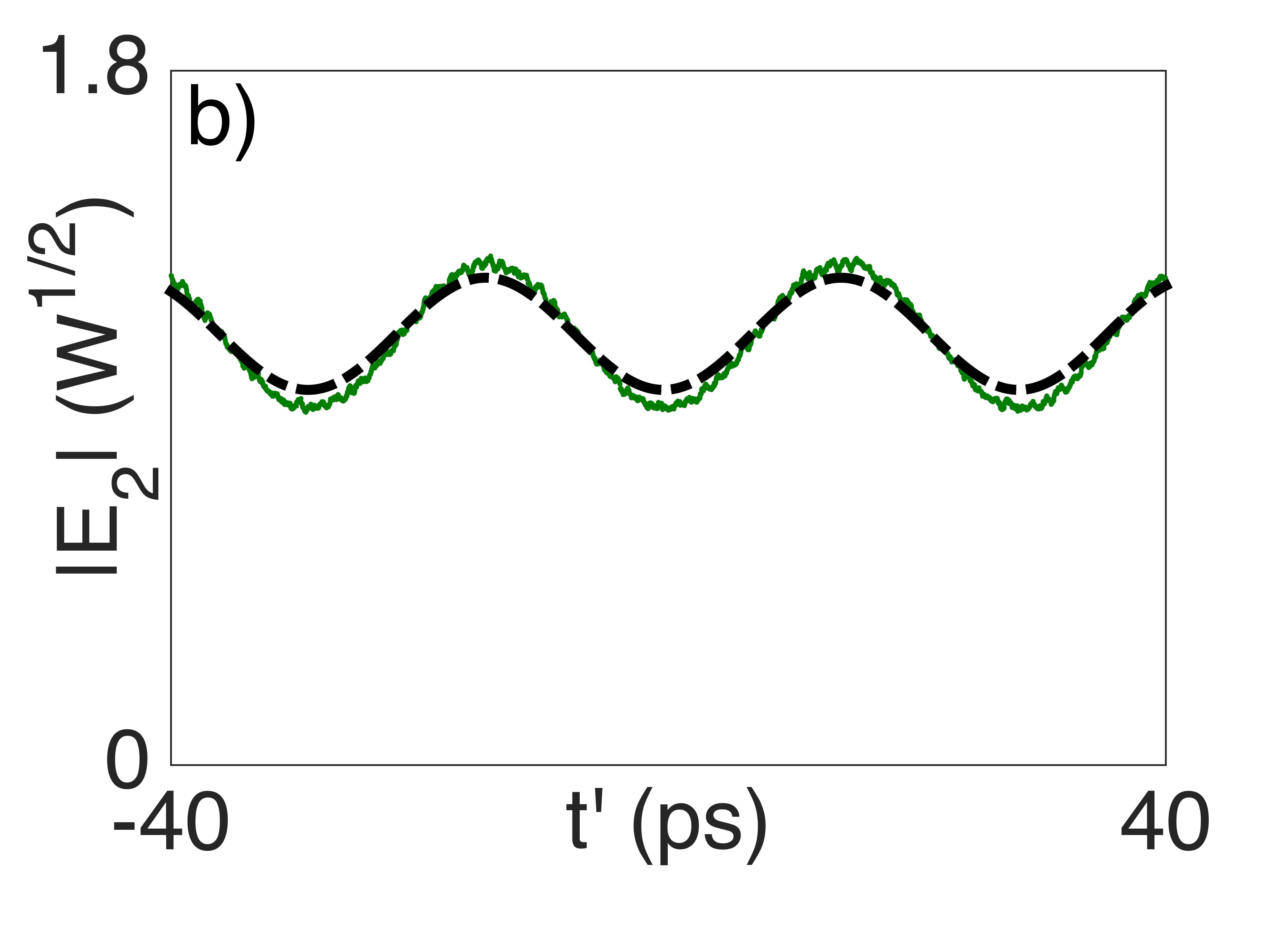}
\includegraphics[width=4cm,height=2.6cm]{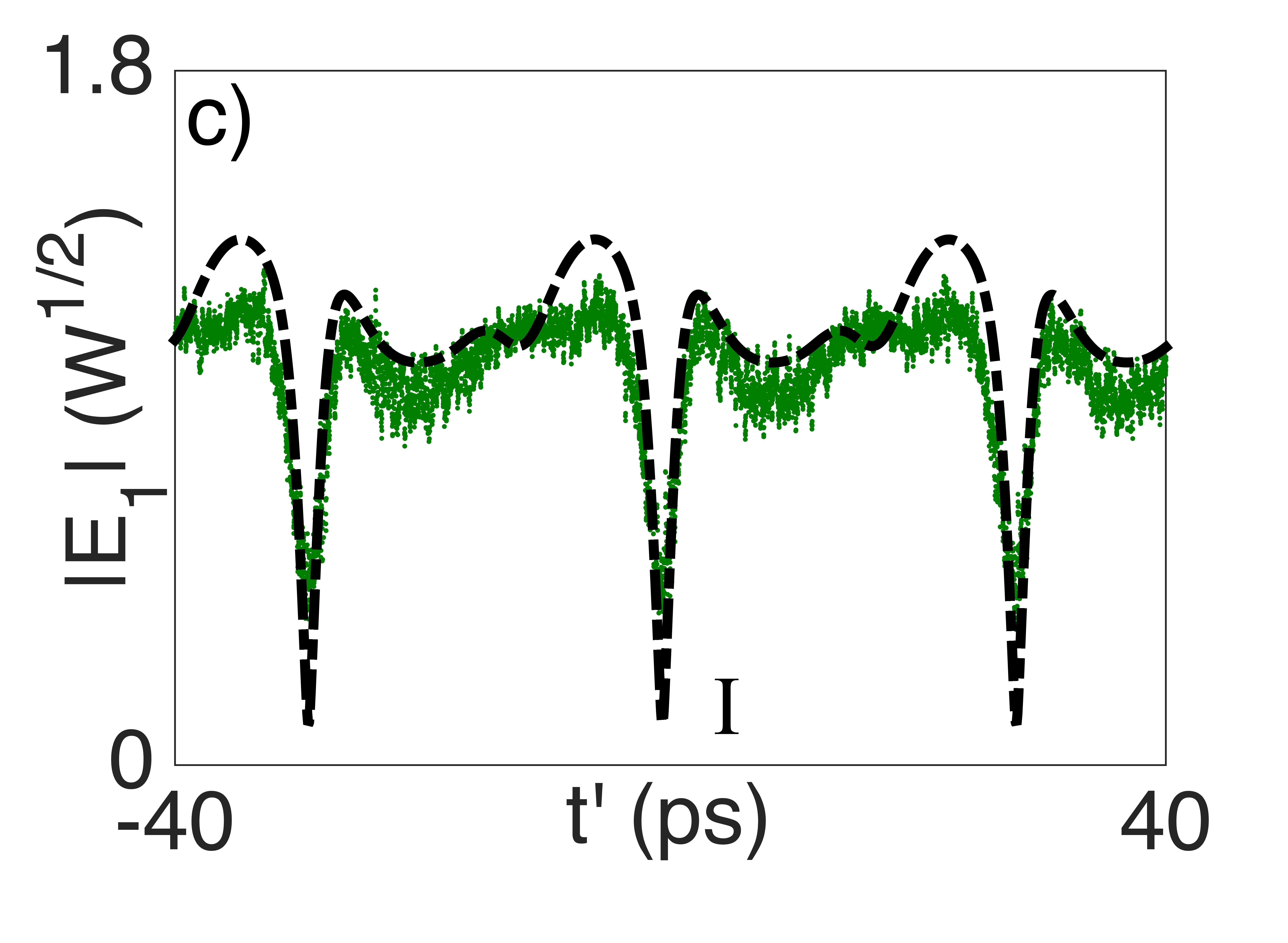}
\includegraphics[width=4cm,height=2.6cm]{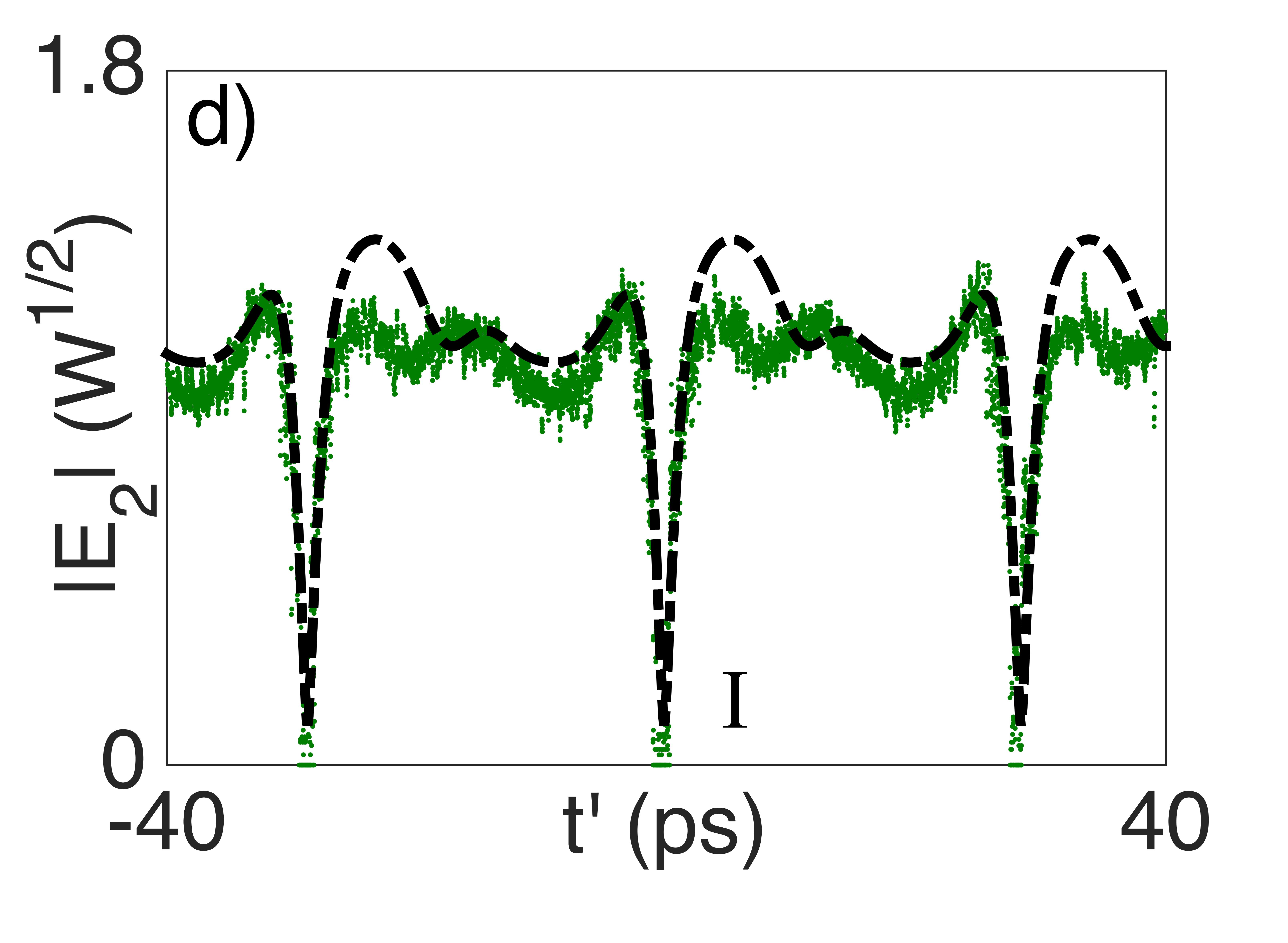}
\includegraphics[width=4cm,height=2.6cm]{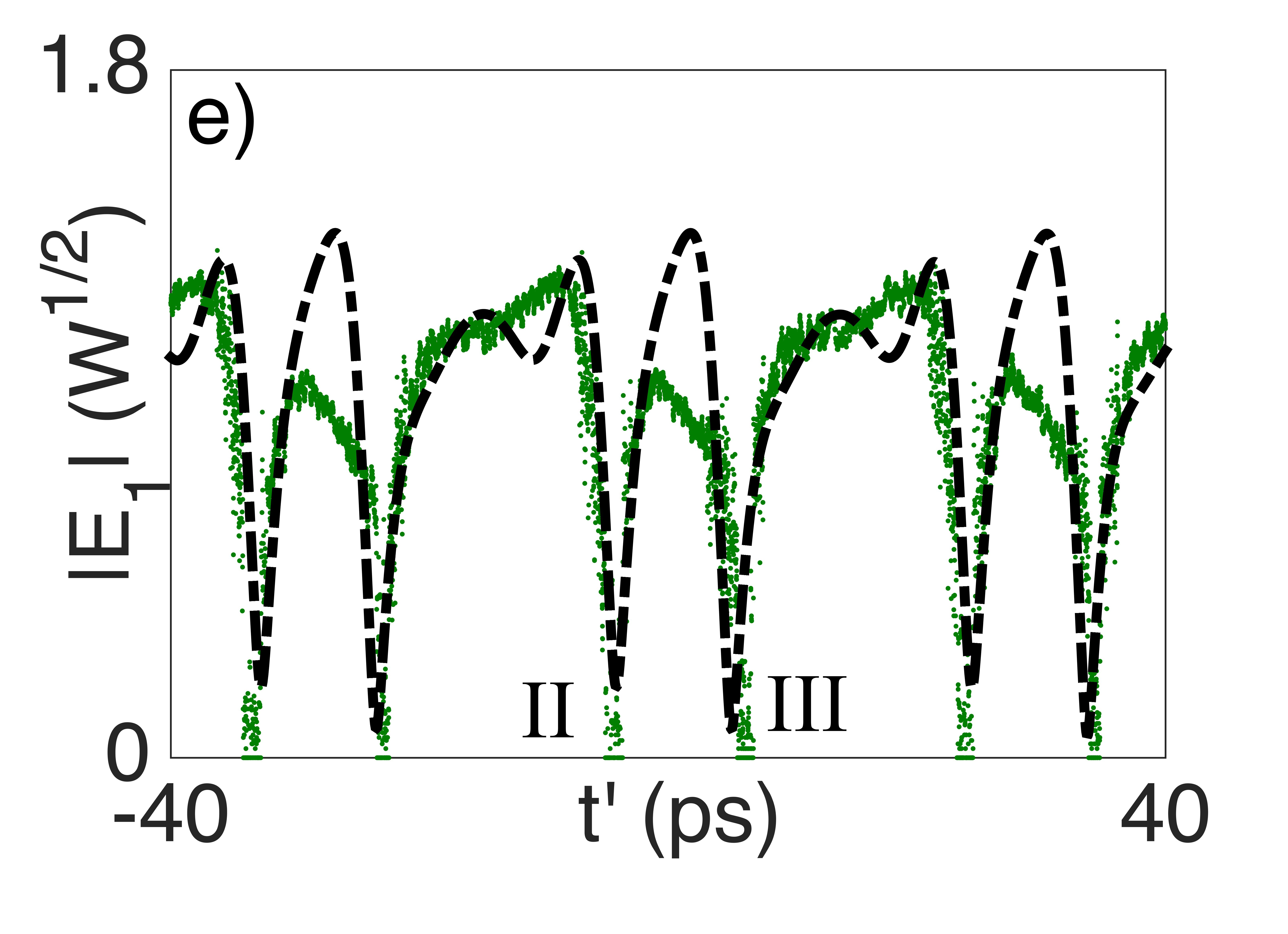}
\includegraphics[width=4cm,height=2.6cm]{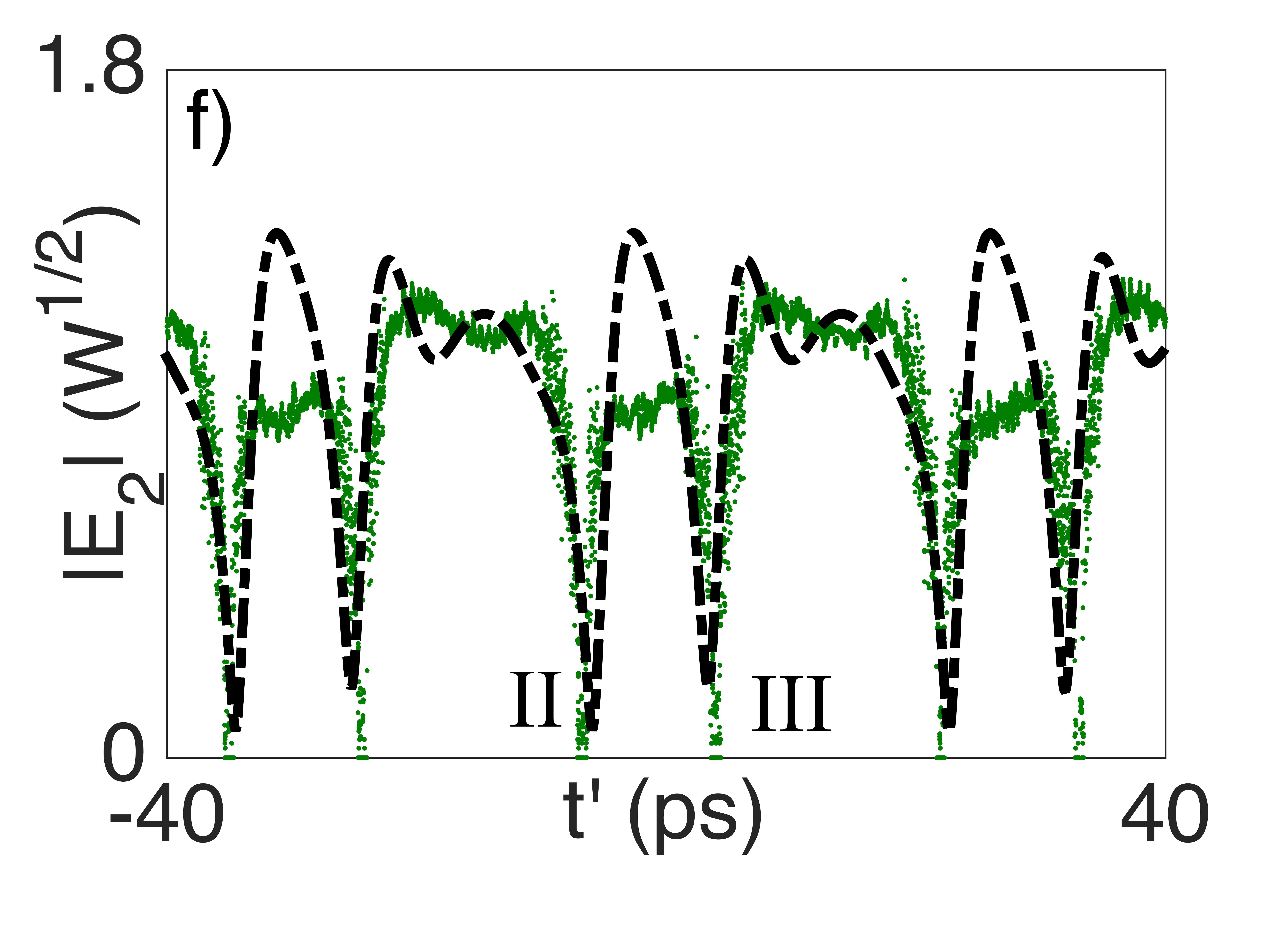}
    \end{center}
     \caption{Comparison between the experimental (green line) and numerical (black dashed line) temporal profiles of 
     $|E_1(t')|$ and $|E_2(t')|$ at the input (a,b), after $3 km$ (c,d) and after $5 km$ (e,f). Numerical 
     results refer to the case reported in figure \ref{fig2}.
    } \label{fig4}
\end{figure}

In addition, Fig. \ref{fig3} presents the comparison between the theoretical and numerical amplitude profiles 
of the two orthogonal polarization waves at the input, after $3 km$ (close to the zero-amplitude point of black rogue wave I),
and after $5km$ (close to the first zero-amplitude points of dark rogue waves II and III). 
As can be seen, an overall excellent agreement is obtained between the analytical solutions and 
the realistic numerical simulations. Slight discrepancies appear due to the non-ideal initial conditions for the 
numerical generation of the \textit{three-sister} dark rogue waves.

 \begin{figure}[h]
\begin{center}
\includegraphics[width=4cm,height=2.5cm]{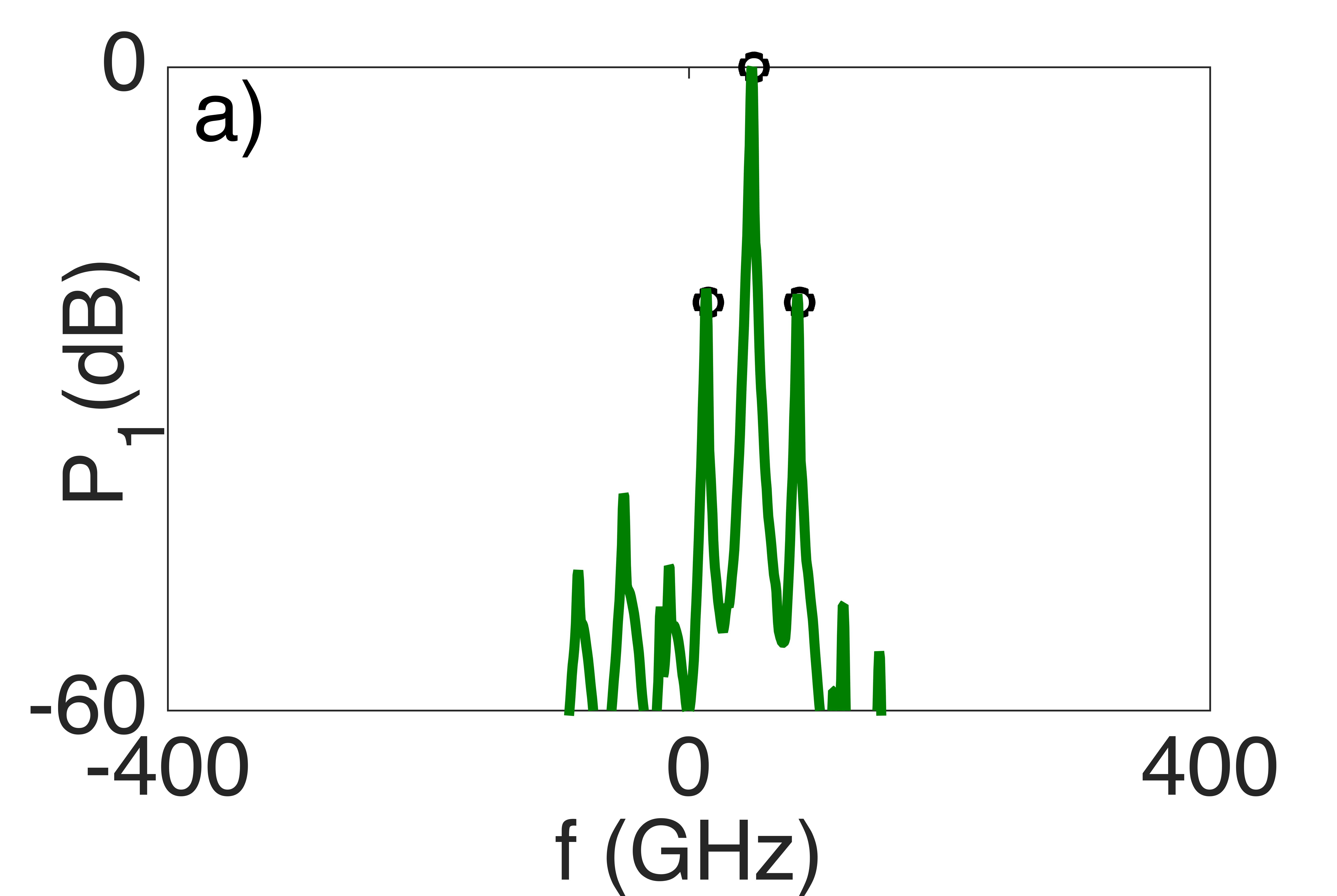}
\includegraphics[width=4cm,height=2.5cm]{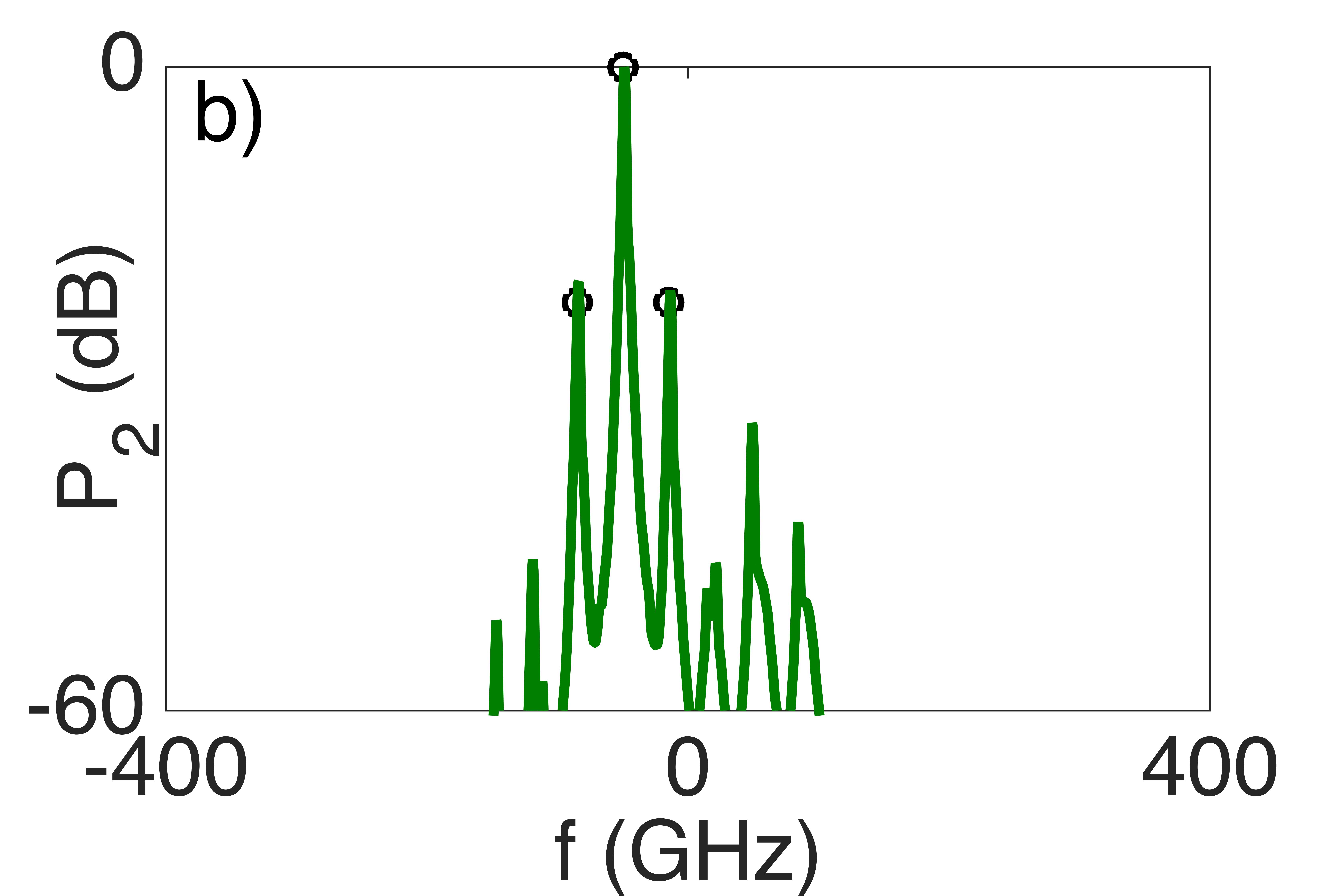}
\includegraphics[width=4cm,height=2.5cm]{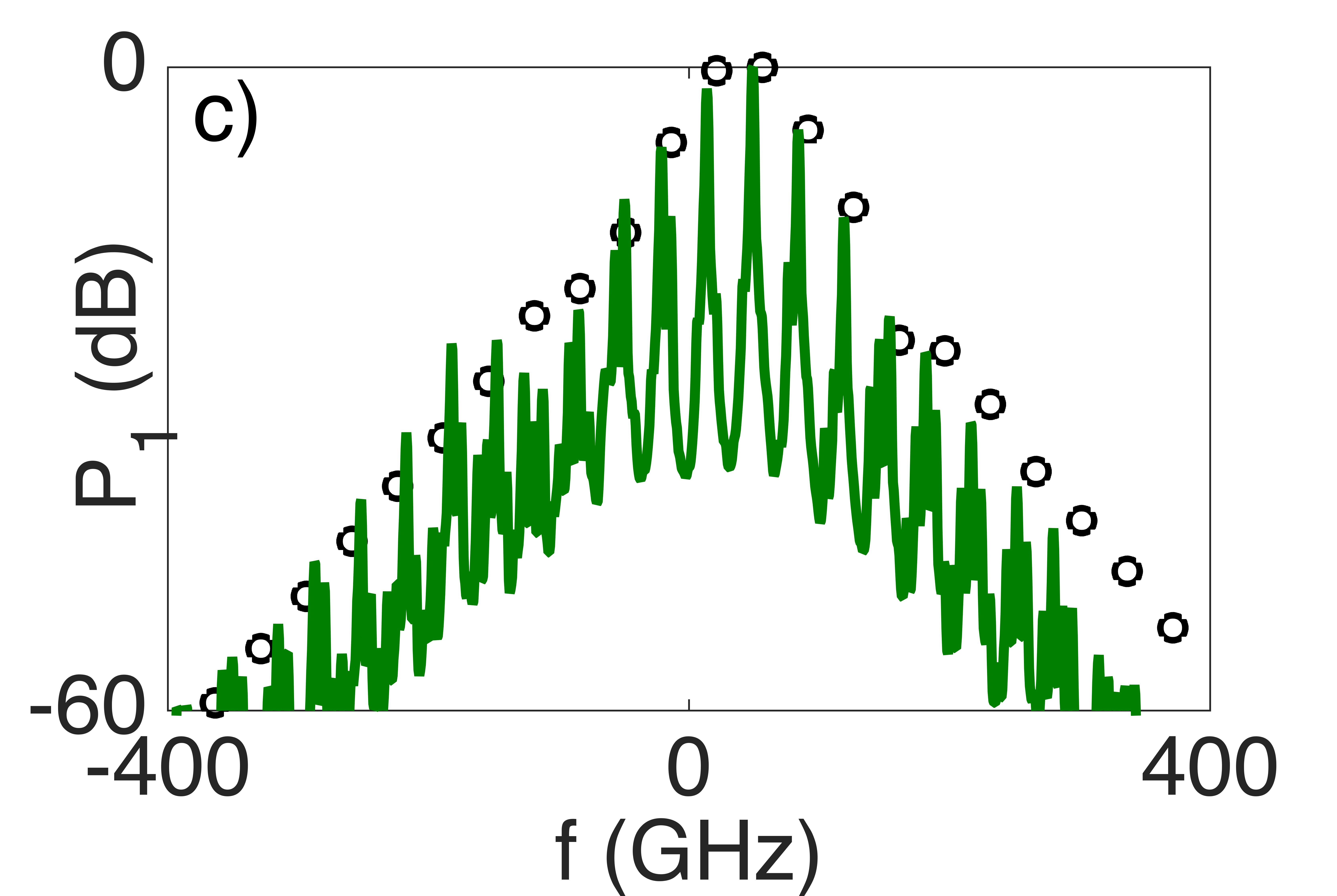}
\includegraphics[width=4cm,height=2.5cm]{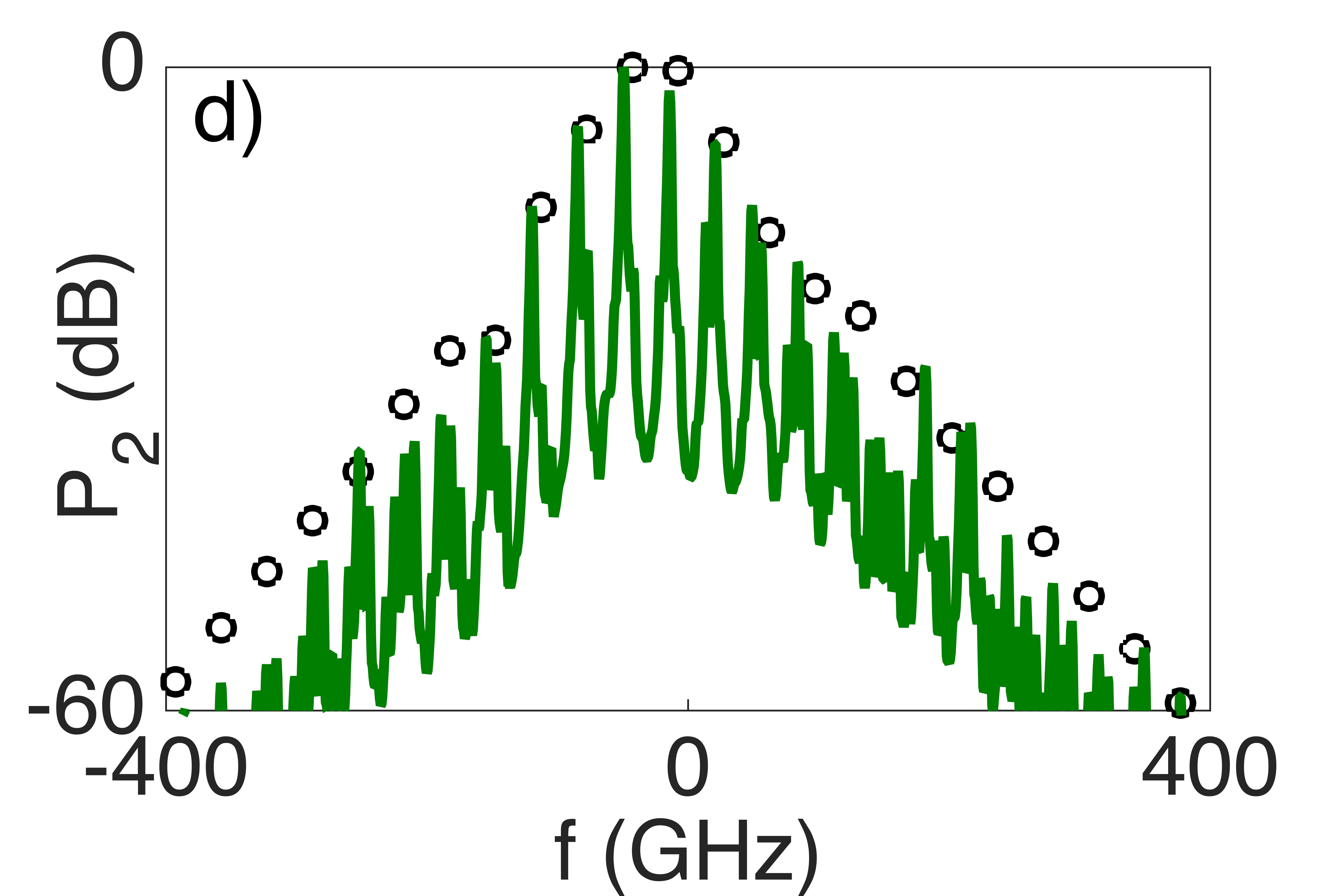}
\includegraphics[width=4cm,height=2.5cm]{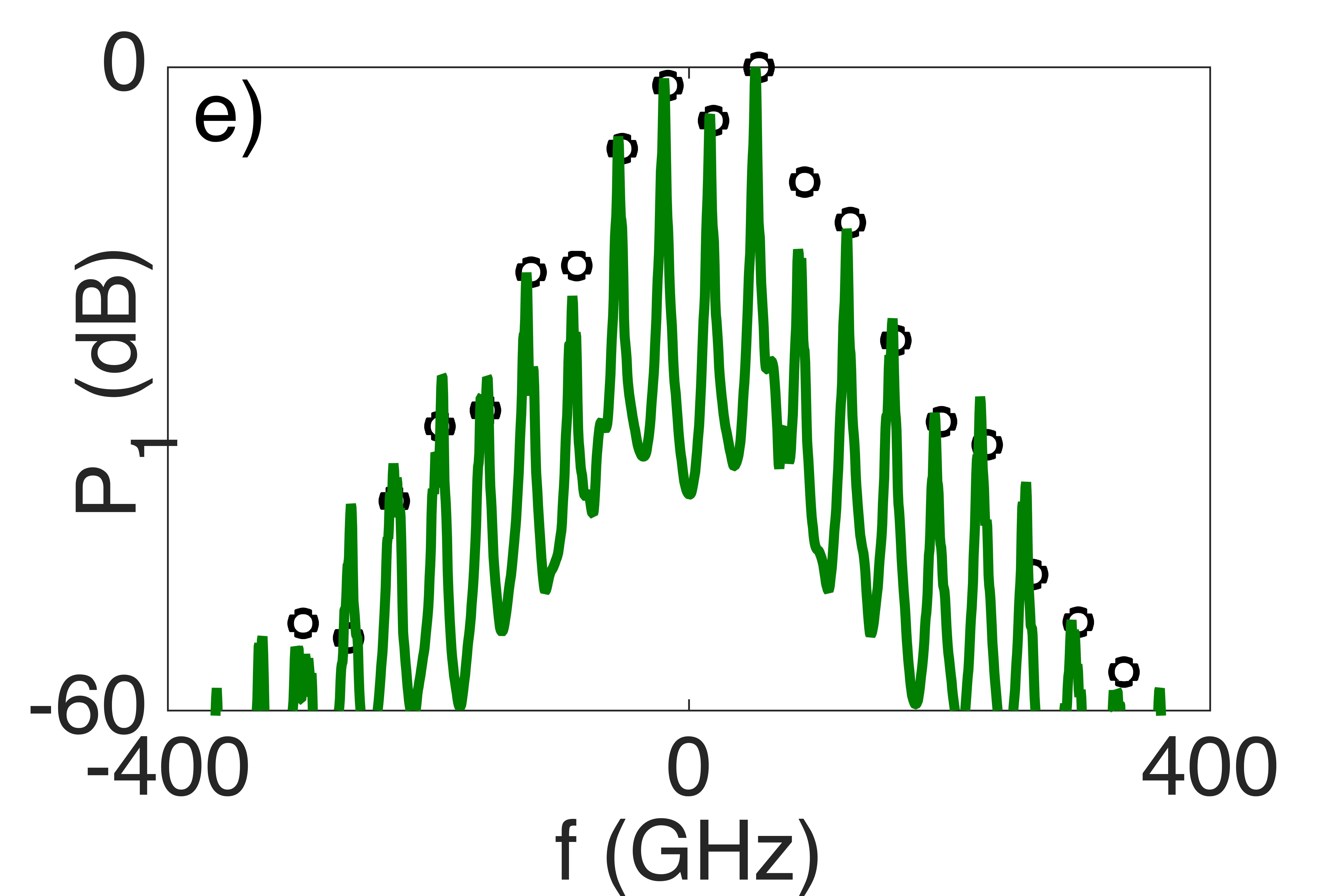}
\includegraphics[width=4cm,height=2.5cm]{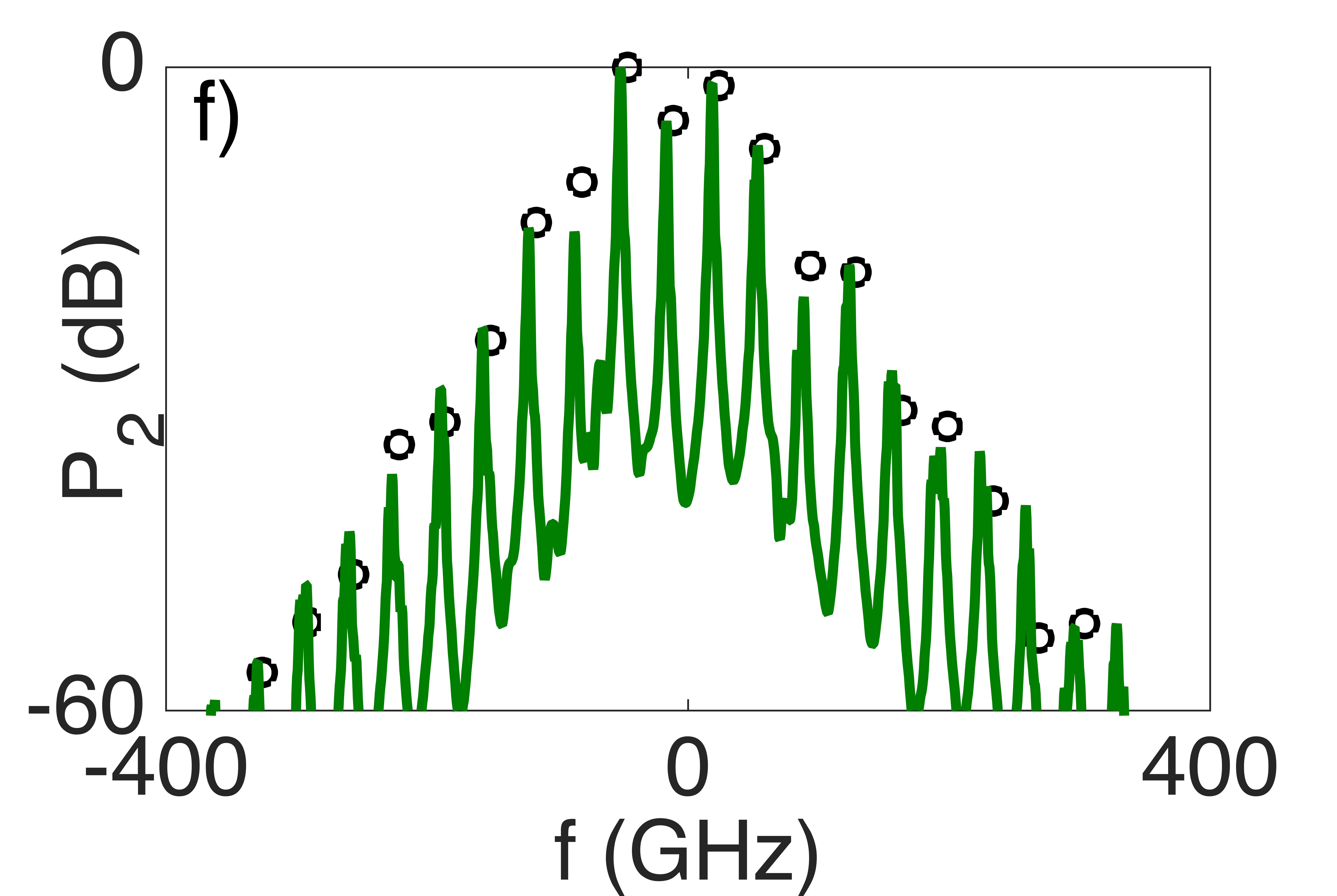}
    \end{center}
     \caption{Comparison between the experimental (green line) and numerical (black circles) spectral profiles of 
     $P_1$ and $P_2$ at the input (a,b), after $3 km$ (c,d) and after $5 km$ (e,f). Numerical 
     results refer to the case reported in figure \ref{fig2}.
    } \label{fig5}
\end{figure}

Quite remarkably, experimental observations are in excellent quantitative agreement, with no adjustable parameters, 
with our theoretical predictions of \textit{three-sister} dark rogue wave generation in optical fiber. Figures \ref{fig4} 
illustrate the measured temporal traces of wave amplitudes emerging from orthogonal polarizations at the fiber input, 
after $3 km$, and after $5 km$, along with their corresponding numerical predictions. 

Additionally, Fig. \ref{fig5} reports both measured and numerically predicted spectral amplitudes  
from the two orthogonal polarization waves at the input, after $3 km$, and after $5 km$, respectively. The measured and numerical traces
show enlarged triangular spectral shapes, which are typical of rogue wave events.
The observed spectral asymmetry also provides a clear signature of the opposite group velocities and nonlinear chirp 
of the temporal rogue structures that have been generated in orthogonal polarizations.

\section{Conclusions}

In brief summary, we have presented an experimental evidence of a group of three coupled dark rogue waves. For doing that, we exploited two orthogonally polarized, nonlinear coupled optical waves, propagating in the defocusing (normal dispersion) regime of a randomly birefringent telecom fiber, as described by the coupled NLSEs or Manakov system.
Although we have used the fiber optics testbed, we would like to emphasize that the higher-order rogue wave solutions that we have described here are of significant interest in the more general multidisciplinary context of rogue waves and extreme events. Namely, such complex rogue waves are expected to significantly influence the heavy-tailed statistics of extreme events, which are observed in many different natural and human complex systems.
Within the scalar NLSE framework, a systematic classification of higher-order rogue wave solutions has led in recent years to their corresponding 
experimental observations, both in hydrodynamics (water tanks) and nonlinear optics (optical fibers). The present work opens the new perspective of high-order rogue wave solutions for
a more general class of multimode coupled wave systems.
Indeed, coupled NLSEs can sustain a vast range of rogue wave structures such as bright-bright, dark-bright, dark-dark solutions. Here, we have focused our attention to the intriguing case of dark-dark rogue wave solutions (more particularly, the second-order solutions or vector dark rogue wave triplets ), which are specific to coupled NLSE systems.
%
%
We would like emphasize that the presently discussed coupled rogue wave prototypes may be considered as the analogous dark optical counterpart of the \textit{three-sister} hydrodynamics rogue waves which have been often mentioned, but to our knowledge without a precise experimental characterization, in past observations of disastrous events in deep water \cite{fitz,niko}.

\section{Acknowledgements}
This research was partially supported by the European Union (Marie Sklodowska-Curie Grant No. 691051), by the National Natural Science Foundation of China (Grants No. 11174050 and No. 11474051),  by the French ``Investissements d'Avenir'' program (project PIA2/ISITE-BFC) and Labex ACTION ANR-11-LABX-0001-01. The project has received funding from the European Research Council (ERC) under the European Union's Horizon 2020 research and innovation programme (grant agreement No.740355).

\end{document}